\documentclass[review,authoryear,12pt]{elsarticle}
\usepackage{graphics}
\usepackage{amsmath}
\allowdisplaybreaks
\usepackage{fancyhdr}
\usepackage{amssymb}
\usepackage{latexsym}
\usepackage{color}
\usepackage{graphicx}
\usepackage{hyperref}
\hypersetup{backref, colorlinks=true}
\usepackage{epsfig}
\usepackage{setspace}
\usepackage{graphicx}
\usepackage{epstopdf}
\usepackage{subfigure}
\usepackage{changes}
\definechangesauthor{red} 
\colorlet{Changes@Color}{red}
\usepackage{float}
\def\rz{\ifmmode{I\hskip -3pt R}
    \else{\hbox{$I\hskip -3pt R$}}\fi}
\def\nz{\ifmmode{I\hskip -3pt N}
    \else{\hbox{$I\hskip -3pt N$}}\fi}
\def\gz{\ifmmode{Z\hskip -4.8pt Z}
    \else{\hbox{$Z\hskip -4.8pt Z$}}\fi}
\def\cz{\ifmmode{C\hskip -4.8pt\vrule height5.8pt\hskip6.3pt}
\else{\hbox{$C\hskip -4.8pt\vrule height6.0pt\hskip6.3pt$}}\fi}
\def\qz{\ifmmode{Q\hskip -5.0pt\vrule height6.0pt depth0pt
   \hskip6pt}
    \else{\hbox{$Q\hskip -5.0pt\vrule height6.0pt depth0pt
   \hskip6pt$}}\fi}

\newcommand{\beq}{\begin{equation}}
\newcommand{\bseqs}{\begin{subequations}}
\newcommand{\eseqs}{\end{subequations}}
\newcommand{\balign}{\begin{aligned}}
\newcommand{\ealign}{\end{aligned}}
\newcommand{\eeq}{\end{equation}}
\newcommand{\beql}{\begin{equation} \label}
\newcommand{\beqs}{\begin{eqnarray}}
\newcommand{\eeqs}{\end{eqnarray}}
\newcommand{\beas}{\begin{eqnarray*}}
\newcommand{\eeas}{\end{eqnarray*}}
\newcommand{\ber}{\begin{array}}
\newcommand{\eer}{\end{array}}
\newcommand{\becs}{\begin{cases}}
\newcommand{\eecs}{\end{cases}}
\newcommand{\bealign}{\begin{aligned}}
 \newcommand{\eealign}{\end{aligned}}

\newcommand{\leftm}{\left[\begin{array}}
\newcommand{\rightm}{\end{array}\right]}

\makeatletter
\def\ps@pprintTitle{%
   \let\@oddhead\@empty
   \let\@evenhead\@empty
   \def\@oddfoot{\reset@font\hfil\thepage\hfil}
   \let\@evenfoot\@oddfoot
}
\makeatother

\journal{}

\begin{document}

\begin{frontmatter}

\title{Surface wrinkling of an elastic block subject to biaxial loading by an energy method}

\author[]{Shengyou Yang\corref{cor1}}
\ead{uhsyyang@gmail.com}
\address{Department of Mechanical Engineering, University of Houston, Houston, TX 77204, USA}
\cortext[cor1]{Corresponding author.}

\date{}

\begin{abstract}
Wrinkles are often observed on the surfaces of compressed soft materials in nature.  In the past few decades, the fascinating surface patterns have been studied extensively by using the linear bifurcation analysis under plane strain. The bifurcation concerns the non-uniqueness solutions, however, it delivers little information about the surface instability before and after the threshold. In this paper, we study surface wrinkling of a finite elastic block of general elastic materials subject to biaxial loading by an energy method. The first and second variations of the strain energy functional are systematically studied, and an eigenvalue problem is proposed whether the second variation is positive definite. We illustrate our analysis by using neo-Hookean materials as an example. Accordingly, we show that the initially flat state has the lowest energy and is stable before the stretches reach the threshold at which the surface wrinkling occurs. We also find that the threshold is independent of the size of the block and coincides with that of the surface instability of an elastic half-space studied by \cite{biot1963surface} with the linear bifurcation analysis. However, the stability region cannot be obtained by using the linear stability analysis. In contrast to the size-independent threshold, the wavelength of surface wrinkling depends on the size of the block. We first show that a two-dimensional rather than a three-dimensional perturbation has lower energy and is more likely to trigger the surface wrinkling in the instability region. The same stretch threshold of a finite block and a half-space could shed light on the relation of surface instabilities between finite and infinite bodies.
\end{abstract}

\begin{keyword}
Surface instability \sep Wrinkling \sep Elastic block \sep Energy minimization, Size effects
\end{keyword}

\end{frontmatter}


\section{Introduction}

Surface instabilities in nature and our daily life have caused the interest of many researchers due to their unique surface topography, high nonlinearity and multi-functional behavior as well as various applications by harnessing or avoiding the surface instabilities. Wrinkles, one typical surface instabilities, are commonly observed when elastic materials especially film/substrate systems are subject to a sufficiently large compression \citep{bowden1998spontaneous, volynskii2000mechanical, yang2010harnessing}. Wrinkles originated from the flat surface have a wave-like geometry with infinitesimal amplitude.

The patterns of wrinkles have been widely used to assemble complex patterns for their potential applications in sensors technology \citep{bowden1998spontaneous}, to produce novel electronic devices \citep{khang2006stretchable}, to fabricate microlens arrays \citep{chan2006fabricating}, to control the adhesion \citep{chan2008surface}, to trigger transformations of phononic band gaps \citep{bertoldi2008mechanically} and an change in the phononic properties \citep{jang2009combining}, and to construct a metrology for measuring mechanical properties of ultra thin polymer films \citep{stafford2004buckling}. In the meanwhile, tremendous theoretical analyses, just name a small sample of the representative works here \citep{groenewold2001wrinkling, shenoy2001pattern, huang2002wrinkling, chen2004herringbone, huang2005kinetic, huang2005nonlinear, jiang2007finite, audoly2008buckling, li2011spontaneous, cao2012wrinkling, chen2012stability, hutchinson2013role, danas2014instability, holland2017instabilities, budday2017wrinkling}, have been carried out to study the wrinkling phenomena of thin films in response to environmental stimuli (e.g., mechanical forces \citep{huang2002wrinkling, huang2005kinetic, huang2005nonlinear, cao2012wrinkling}, temperature \citep{chen2004herringbone}, magnetic field \citep{danas2014instability}, and van der Waals interactions \citep{shenoy2001pattern, li2011spontaneous, chen2012stability}). 

In addition to the extensive study of wrinkles on film/substrate systems in the past few decades, the mechanism of surface instability on homogeneous elastic block is fundamental and is of highly physical and mathematical interests. The root of surface instability can be found in the pioneering work of surface instability of rubber in compression by \cite{biot1963surface}. At a critical compressive strain $0.46$ for plane strain, Biot showed that the surface of a half-space of a homogeneous incompressible neo-Hookean material became unstable. Biot pointed out that the critical strain was independent of the elastic modulus, moreover, the wavelength of the surface mode was undetermined and could be arbitrarily short or long because there was no physical length quantity in a half-space problem. Subsequently, a mount of theoretical analyses \citep{levinson1968stability, nowinski1969surface, usmani1974surface, chen2018surface} have been carried out to study surface instabilities of a half-space of elastic materials. It is worth mentioning that the work \citep{biot1963surface, nowinski1969surface, usmani1974surface} are actually based on the linear stability analysis that merely solves the incremental equilibrium equations. The incremental solution cannot give the complete set of stability conditions of an elastic half-space. 

Biot's theoretical prediction of surface instability remained unchallenged for a long time until \cite{gent1999surface} found its apparent disagreement with their experimental result. To verify Biot's theoretical prediction of surface instability, they performed an experiment on the bending of a rubber block.  Since the occurrence of unidirectional compression on the inner surface of a rubber block subject to simple bending, a similar surface instability would be expected at a critical degree of bending when the surface compression was about $0.46$ from Biot's prediction. However, their experiments showed that sharp creases occurred on the inner surface at a critical compressive strain $0.35$ that was less than predicted, only about one-half as severe. This discrepancy was not known at that moment until \cite{hohlfeld2011unfolding} proposed that the formation of a crease was a new type of instability. They illustrated creases both by numerical simulations and experiments of a bent slab of an incompressible elastic material. \cite{hong2009formation} obtained the same critical strain of the crease formation by comparing the elastic energy in a creased body and that in a homogeneous body in their finite-element analysis (FEA). Other numerical approaches \citep{wong2010surface, tallinen2013surface} also obtained the critical strains for the onset of creases on the free surfaces. \cite{cao2012wrinkles} showed, using Koiter's initial post-buckling analysis \citep{koiter1945stability, van2009wt}, that wrinkling was extremely unstable and highly imperfection-sensitive. They found that the initial imperfections highly reduced the critical compressive strain for the onset of surface instability. 

A brief literature review shows that theoretical analysis of surface instability can be traced back to the pioneering work by \cite{biot1963surface}, which investigated the surface instability of an elastic half-space by using the linear stability analysis. However, Biot's prediction is challenged by the experiments \citep{gent1999surface} regarding the occurrence of creases prior to wrinkles. Up to now, no experiments report the observation of Biot's smooth wrinkles on the surface of homogeneous elastic materials under compression. Owing to a wide range of important applications, the phenomena of surface instabilities have attracted the attention of many researchers around the world. Although significant progress has been made on the analysis of surface instabilities over the past several decades, there still exists a large number of interesting problems that are of physical and mathematical interests. In this paper, we focus on the formation of wrinkles rather than creases by revisiting Biot's pioneering work. 

The purpose of this paper is threefold. First, we study the wrinkles on finite elastic blocks rather than on elastic half-spaces to explore the effects of all the boundary conditions on the formation of wrinkles. Second, the stability criterion used in this paper is the principle of minimum energy rather than the linear bifurcation analysis used in most of the previous works. Third, a three-dimensional perturbation rather than a two-dimensional perturbation is considered in order to explore more diverse surface patterns, especially the possibility of some patterns that can trigger the surface wrinkling at relatively low strains.

A half-space has only one surface that divides the three-dimensional space into two parts. As an ideal mathematical model, the half-space is commonly used to investigate the mechanical behavior of large solids subject to environmental stimuli. The domain occupied by the half-space can be extended to infinite length. However, most of the specimens in experiments are finite large rather than infinitely large. The natural question is to ask whether the size and the boundary conditions of large solids, in reality, have some effects on the threshold on the surface wrinkling. \cite{yang2017wrinkle} studied the boundary conditions on the surface wrinkling of an inhomogeneous elastic block with graded stiﬀness by using the linear stability analysis under plane strain. A more general three-dimensional surface wrinkling based on the energy method has not yet been reported.

The linear stability analysis is widely used to investigate the surface wrinkling in the previous work. However, differences between stability and bifurcation do exist \citep{ericksen1956implications, hill1957on}. The linear bifurcation analysis investigates the solution of the incremental equilibrium equations. The existence of an incremental solution is just a necessary condition for the existence of a bifurcation point. It delivers little information about the stability before or after the bifurcation points \citep{chen2001singularity, chen2018surface}. In contrast, the energy method based on the principle of minimum energy has rigorous physical and mathematical bases, which is used in this paper to investigate the surface instability of an elastic block subject to biaxial loading.

This paper is organized as follows. In Section \ref{section2}, we establish an elastic block problem with the constraint of incompressibility and the kinematic boundary conditions.  Accordingly, Section \ref{section3} is devoted to the stability criterion of an elastic block at finite deformation where we use the principle of minimum energy. We compute the first and second variations of the strain energy functional. Subsequently, we extreme the second variation in a compact set and then construct an eigenvalue problem whose eigenvalue is exactly the value of the second variation at the corresponding eigenfunction. The requirement of a positive semi-definite second variation at equilibrium is converted into the condition for all non-negative eigenvalues, including the lowest eigenvalue that is exactly the lower bound of the second variation. In Section \ref{section4}, we carry out our analysis by using neo-Hookean materials. Solution of the boundary-value problem gives a trivial solution corresponding to the homogeneous deformation. Stability of the homogeneous deformation is examined by solving the eigenvalue problem with double Fourier series, which finally becomes the discussion of the characteristic equation. In Section \ref{section5}, we discuss the stability conditions by examining the sign of the eigenvalue. We give the stability and instability regions in the principal stretch plane and discuss the wrinkling patterns in the instability region. Conclusions are given in Section \ref{conclusion}.

\section{Formulation} \label{section2}

\begin{figure}[h] 
\centering
\subfigure[]{%
        \includegraphics[width=2in]{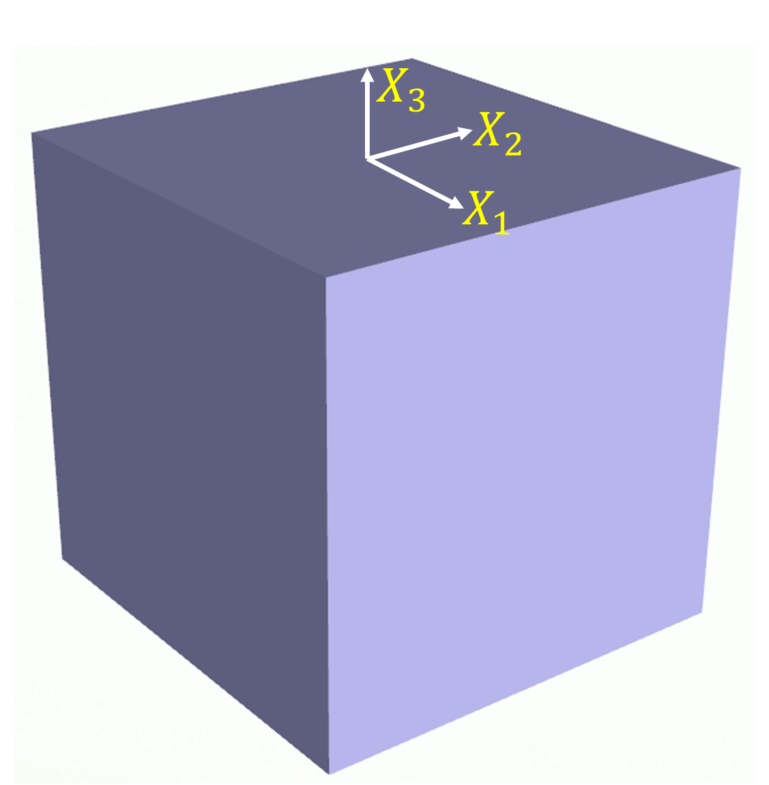}
       \label{fig1a} } 
\subfigure[]{%
        \includegraphics[width=2in]{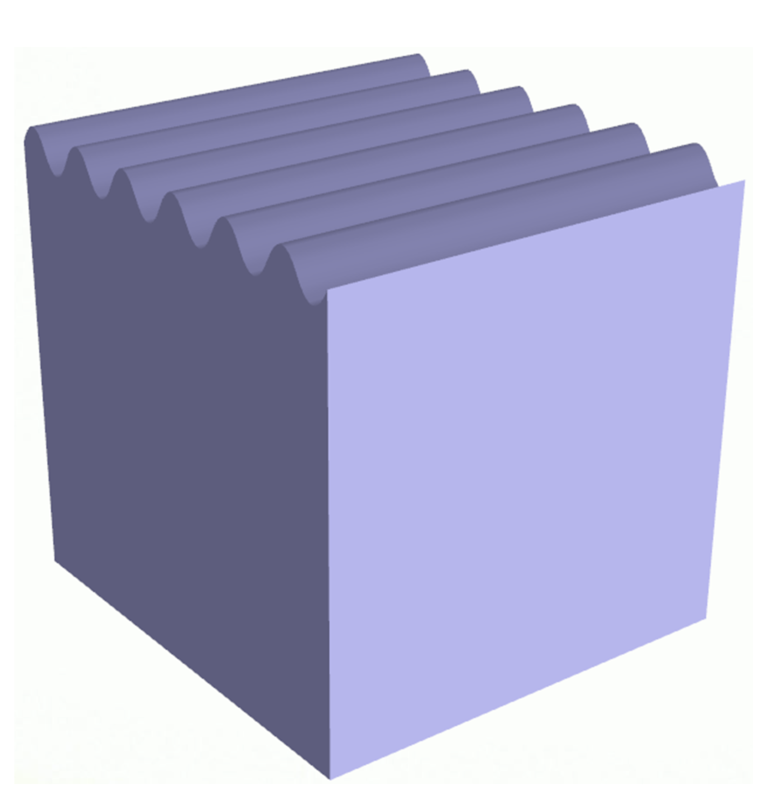}
       \label{fig1b} }
       \subfigure[]{%
        \includegraphics[width=2in]{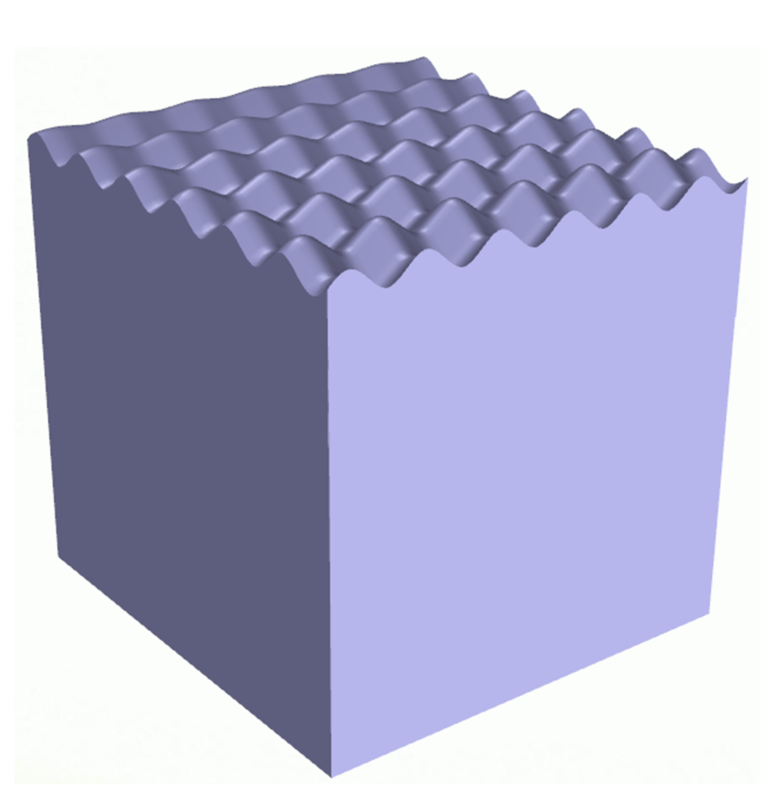}
       \label{fig1c} }
\caption{ Schematic of an elastic block subject to biaxial loading on the four lateral surfaces: (a) undeformed block with a flat upper surface; (b) two-dimensional surface wrinkling in the $X_1-X_3$ plane, i.e., the periodical wrinkles are aligned with the $X_2$ direction; (c) three-dimensional surface wrinkling on the top surface of the elastic block.}
\label{fig1}
\end{figure}

Consider an elastic body that occupies, in a reference configuration with a suitably chosen right-handed, orthonormal basis $\{ {\bf e}_1, {\bf e}_2, {\bf e}_3\}$, a rectangular block represent by
\begin{equation}\label{RDomain}
\mathcal{B}=\{\mathbf{X}\in\mathbb{R}^3: |X_1|\le l_1, |X_2|\le l_2, -l_3 \le X_3 \le 0\},
\end{equation}
where $l_i$, $i=1,2,3$, are given positive constants and $X_i = {\bf X} \cdot {\bf e}_i$ is the coordinate in the ${\bf e}_i-$direction. Since the elastic block $\mathcal{B}$ consists of six surfaces, its boundary $\partial\mathcal{B}$ can be divided into six parts (see Figure \ref{fig1}). The choice of this geometry is motivated by the need to formulate a physical reasonable boundary-value problem in studying instability of elastic bodies with free surfaces. 

The body may undergo a deformation which is expressed by a smooth function $\mathbf{x}:\mathcal{B}\to\mathbb{R}^3$. The elastic body is assumed to be incompressible, which requires that
\begin{equation}\label{RD_G}
\det\mathbf{F}=1,
\end{equation}
where $\mathbf{F}=\triangledown\mathbf{x}$ is the deformation gradient. We shall consider the physical situation where the deformation is controlled by a loading device that stretches or compresses the body in the direction of coordinate axes. Precisely, the deformation is also required to satisfy the following kinematical boundary conditions that prescribe the normal displacement of the four sides surfaces, as well as the displacement at the bottom surface:
\begin{equation}\label{RB_C_1}
x_{\alpha}=\lambda_{\alpha}X_{\alpha} \quad \textrm{at} \  X_{\alpha}=\pm l_{\alpha},\quad \alpha=1,2\ \textrm{(no sum)}
\end{equation}
and
\begin{equation}\label{RB_C_2}
\mathbf{x(X)}=\lambda_1X_1\mathbf{e_1}+\lambda_2X_2\mathbf{e_2}+\frac{1}{\lambda_1\lambda_2}X_3\mathbf{e_3}  \quad \textrm{at} \  X_3=- l_3,
\end{equation}
where $\lambda_1$ and $\lambda_2$ are two stretches. Physically, the boundary conditions $\eqref{RB_C_1}$ can be implemented by four rigid smooth plates that are in contact with the side surfaces of $\mathcal{B}$ and that allow surface sliding \citep{chen2012stability, yang2017wrinkle, yang2017revisiting}. In contrast, the boundary condition $\eqref{RB_C_2}$ donates a displacement-prescribed bottom surface \citep{yang2017wrinkle}. The present boundary conditions provide a physical setting that facility, among other things, such periodical solutions.

\section{Energy stability criterion}\label{section3}

By the principle of minimum energy, deformation $\mathbf{x}$ is stable if a properly defined potential energy at $\mathbf{x}$ is not greater than the potential energies at other deformations in an appropriate neighborhood of $\mathbf{x}$. For the present problem, an elastic block with a traction-free upper surface and boundary conditions \eqref{RB_C_1} and \eqref{RB_C_2}, the potential energy consists of only the elastic energy stored in the deformed body. Such a potential energy is defined by
\begin{equation}\label{Rsub_Energy}
E[\mathbf{x}]=\int_{\mathcal{B}}W(\mathbf{F})dV,
\end{equation}
where $W$ is the strain-energy function of the elastic body. The deformation $\mathbf{x}$ is said to be stable if 
\begin{equation}\label{RE<E}
E[\mathbf{x}]\le E[\tilde{\mathbf{x}}]
\end{equation}
for each $\tilde{\mathbf{x}}$ that satisfies the incompressibility constraint $\eqref{RD_G}$ and the boundary conditions $\eqref{RB_C_1}$ and $\eqref{RB_C_2}$, and that belongs to a neighborhood of $\mathbf{x}$ with respect to a properly chosen topology.
\subsection{First and second variation conditions}\label{sec3.1}

The inequality $\eqref{RE<E}$ leads to a constrained minimization problem with nonlinear constraint $\eqref{RD_G}$. Follow the approach of \cite{fosdick1986minimization}, we minimize the potential
energy directly in the set of the kinematically admissible deformations and seek minima of $E$ in a parameterized subset of all admissible deformations $\mathbf{\tilde{x}}$. Let $\mathbf{\tilde{x}}(\mathbf{X},\epsilon)$ be a one-parameter family of functions that satisfy $\eqref{RD_G}$-$\eqref{RB_C_2}$ for all $\epsilon$, and that satisfy
\begin{equation}\label{Rx_hat}
\mathbf{\tilde{x}}(\mathbf{X},0)=\mathbf{x}(\mathbf{X}).
\end{equation}
Inequality $\eqref{RE<E}$ then implies that
\begin{equation}\label{RF_V_2}
\dot{E}=0
\end{equation}
and
\begin{equation}\label{RS_V_2}
\ddot{E}\ge0
\end{equation}
for all possible choices of $\mathbf{\tilde{x}}(\mathbf{X},\epsilon)$, where
\begin{equation}\label{3.10}
\dot{E}=\left.\frac{\partial{E[\mathbf{\tilde{x}}(\mathbf{X},\epsilon)]}}{\partial\epsilon}\right|_{\epsilon=0},\ \ddot{E}=\left.\frac{\partial^2{E[\mathbf{\tilde{x}}(\mathbf{X},\epsilon)]}}{\partial\epsilon^2}\right|_{\epsilon=0}.
\end{equation}
Equation $\eqref{RF_V_2}$ is also called the first variation condition, whose explicit form here is
\begin{equation}\label{3.11}
\int_{\mathcal{B}}\frac{\partial W}{\partial{\mathbf{F}}}{\cdot}\triangledown{\mathbf{u}}\,dV=0,
\end{equation}
where $\mathbf{u}$ is the first variation of $\tilde{\mathbf{x}}$, defined by
\begin{equation}\label{Ru}
\mathbf{u}=\left.\frac{\partial{\tilde{\mathbf{x}}(\mathbf{X},\epsilon)}}{\partial\epsilon}\right|_{\epsilon=0}.
\end{equation}
Since $\tilde{\mathbf{x}}(\mathbf{X},\epsilon)$ satisfies $\eqref{RD_G}$-$\eqref{RB_C_2}$ for all $\epsilon$, function $\mathbf{u}$ must satisfy the constraint
\begin{equation}\label{RuF}
\triangledown{\mathbf{u}}\cdot \mathbf{F}^{-T}=0
\end{equation}
and the boundary conditions
\begin{equation}\label{Ru_B_C_1}
u_\alpha=0 \quad \textrm{at} \  X_\alpha=\pm l_\alpha, \quad \alpha=1,2
\end{equation}
and
\begin{equation}\label{Ru_B_C_2}
\mathbf{u(X)=0}   \quad \textrm{at} \  X_3=- l_3.
\end{equation}
Using the divergence theorem in $\eqref{3.11}$, we obtain
\begin{align}
0&=\int_{\mathcal{B}}\frac{\partial W}{\partial{\mathbf{F}}}{\cdot}\triangledown{\mathbf{u}}\,dV=\int_{\mathcal{B}}\left\{\mathrm{Div}\left[\left(\frac{\partial W}{\partial{\mathbf{F}}}\right)^T[\mathbf{u}]\right]-\mathbf{u}\cdot\mathrm{Div}\frac{\partial W}{\partial{\mathbf{F}}} \right\}dV\nonumber\\
&=\int_{\partial\mathcal{B}}\mathbf{N}\cdot\left(\frac{\partial W}{\partial{\mathbf{F}}}\right)^T[\mathbf{u}]dA-\int_{\mathcal{B}}\mathbf{u}\cdot\mathrm{Div}\frac{\partial W}{\partial{\mathbf{F}}} dV\nonumber\\
&=\int_{\partial\mathcal{B}}\mathbf{u}\cdot\frac{\partial W}{\partial{\mathbf{F}}}[\mathbf{N}]dA-\int_{\mathcal{B}}\mathbf{u}\cdot\mathrm{Div}\frac{\partial W}{\partial{\mathbf{F}}} dV,
\end{align}
where `$\mathrm{Div}$' denotes the divergence operator with respect to $X$ and $\mathbf{N}$ is the unit outward normal to the surface $\partial\mathcal{B}$ in the reference configuration. Since $\mathbf{\tilde x}{(\mathbf{X},0)}$ is a minimizer of the potential energy function $E$, the last expression in the above equation must vanish at $\epsilon=0$. Considering $\eqref{RuF}$-$\eqref{Ru_B_C_2}$ and using the argument of \cite{fosdick1986minimization}, there exists a smooth function $p$: $\mathcal{B}$ $\to$ $\mathbb{R}$.\\
\indent Equation $\eqref{3.11}$ holds for all $\mathbf{u}$ that satisfy $\eqref{RuF}$-$\eqref{Ru_B_C_2}$ only if
\begin{equation}\label{RE_L_Equation}
\textrm{Div}\left(\frac{\partial W}{\partial{\mathbf{F}}}-p\mathbf{F}^{-T}\right)=\mathbf{0}\quad \textrm{in} \  \mathcal{B},
\end{equation}
\begin{equation}\label{RNatural_B_C}
\left(\frac{\partial W}{\partial{\mathbf{F}}}-p\mathbf{F}^{-T}\right)\mathbf{e}_{\alpha}=z_{\alpha}\mathbf{e}_{\alpha}\quad \textrm{at} \  X_\alpha=\pm l_\alpha, \quad \alpha=1,2\ (\textrm{no sum}),\ \textrm{and}
\end{equation}
\begin{equation}\label{RNatural_B_C_2}
\left(\frac{\partial W}{\partial{\mathbf{F}}}-p\mathbf{F}^{-T}\right){\mathbf{e}_{3}}=\mathbf{0} \quad \textrm{at}\ X_3=0,
\end{equation}
where $p$ is the hydrostatic pressure required by the incompressibility constraint and $z_{\alpha}$ is the normal stress on the side surface. Equation $\eqref{RE_L_Equation}$ is the equilibrium equation, $\eqref{RNatural_B_C}$ are the traction boundary conditions on the side surfaces, and $\eqref{RNatural_B_C_2}$ is the traction boundary condition on the free surface $X_3=0$. These equations, along with the constraint $\eqref{RD_G}$ and the kinematical boundary conditions $\eqref{RB_C_1}$-$\eqref{RB_C_2}$, form a boundary-value problem whose solutions, for prescribed values of $\lambda_1$ and $\lambda_2$, given possible equilibrium deformations, including all stable deformations. \\
\indent The so-called second variation condition $\eqref{RS_V_2}$, which likes $\eqref{RF_V_2}$ is a necessary condition for the deformation $\mathbf{x}$ to be stable, can be written explicitly as
\begin{align}\label{3.19}
\int_{\mathcal{B}}\left\{\frac{\partial W}{\partial{\mathbf{F}}}{\cdot}\triangledown{\mathbf{v}}+\triangledown\mathbf{u}\cdot\frac{\partial^2 W}{\partial\mathbf{F^2}}[\triangledown{\mathbf{u}}]\right\}\,dV\ge 0,
\end{align}
where $\mathbf{v}$ is the second variation of $\mathbf{x}$, defined by
\begin{equation}\label{Rv}
\mathbf{v}=\left.\frac{\partial^2{\tilde{\mathbf{x}}(\mathbf{X},\epsilon)}}{\partial\epsilon^2}\right|_{\epsilon=0}.
\end{equation}
It follows from $\eqref{RD_G}$-$\eqref{RB_C_2}$ that $\mathbf{v}$ satisfies the constraint condition, namely
\begin{equation}\label{RvF}
\triangledown\mathbf{v}\cdot\mathbf{F}^{-T}-\textrm{tr}(\mathbf{F}^{-1}\triangledown\mathbf u)^2=0,
\end{equation}
and the boundary conditions
\begin{equation}\label{Rv_B_C_1}
v_\alpha=0 \quad \textrm{at} \  X_\alpha=\pm l_\alpha, \quad \alpha=1,2
\end{equation}
and
\begin{equation}\label{Rv_B_C_2}
\mathbf{v(X)=0}   \quad \textrm{at} \  X_3=- l_3.
\end{equation}
By using $\eqref{RE_L_Equation}$-$\eqref{RNatural_B_C_2}$ and $\eqref{Rv}$-$\eqref{Rv_B_C_2}$, we find that $\eqref{3.19}$ becomes
\begin{align}\label{3.24}
\int_{\mathcal{B}}\left\{p\, \textrm{tr}\left(\mathbf{ F}^{-1}\triangledown\mathbf{u}\right)^2+\triangledown\mathbf{u}\cdot\frac{\partial^2{W}}{\partial\mathbf{F}^2}[\triangledown\mathbf{u}]\right\}dV \ge 0.
\end{align}

A necessary condition for deformation $\mathbf{x}$ to be stable is then that inequality $\eqref{3.24}$ holds for all $\mathbf{u}$ that satisfy $\eqref{RuF}$-$\eqref{Ru_B_C_2}$. We shall solve this quadratic integral inequality by seeking minima of the integral in a compact subset of the class of admissible functions $\mathbf{u}$. 

\subsection{Eigenvalue problem associated with the stability}\label{sec3.2}

The eigenvalue approach has been recently used to study the surface instability of elastic half-spaces \citep{chen2018surface}. It is observed that $\eqref{3.24}$ holds for all $\mathbf{u}$ that satisfy $\eqref{RuF}$-$\eqref{Ru_B_C_2}$ if and only if it holds for all $\mathbf{u}$ that satisfy $\eqref{RuF}$-$\eqref{Ru_B_C_2}$ and the following normalization condition
\begin{equation}\label{Rnormal}
\int_{\mathcal{B}}\left|\triangledown\mathbf{u}\right|^2dV=1.
\end{equation}
\indent The integral in $\eqref{Rnormal}$ is bounded below on the set of all functions $\mathbf{u}$ that satisfy $\eqref{RuF}$-$\eqref{Ru_B_C_2}$ and $\eqref{Rnormal}$. It then follows that there exists $\mu\in\mathbb{R}$ such that
\begin{equation}\label{R>mu}
 \int_{\mathcal{B}}\left\{p\, \textrm{tr}\left(\mathbf{ F}^{-1}\triangledown\mathbf{u}\right)^2+\triangledown\mathbf{u}\cdot\frac{\partial^2{W}}{\partial\mathbf{F}^2}[\triangledown\mathbf{u}]\right\}dV\ge \mu
\end{equation}
for all $\mathbf{u}$ that satisfy $\eqref{RuF}$-$\eqref{Ru_B_C_2}$ and $\eqref{Rnormal}$, with the equality in $\eqref{R>mu}$ holding for some $\mathbf{u}$. This particular $\mathbf{u}$ can be found by minimizing the integral subject to $\eqref{RuF}$-$\eqref{Ru_B_C_2}$ and $\eqref{Rnormal}$. Thus we have a constrained minimization problem. To find the particular ${\bf u}$ satisfying the equality in \eqref{R>mu}, we take the first variation of this constrained minimization problem and we have the following boundary-value problem:
\begin{equation}\label{3.27}
\textrm{Div}\left\{p\mathbf{F}^{-T}\triangledown\mathbf{u}^T\mathbf{F}^{-T}+\frac{\partial^2W}{\partial\mathbf{F}^2}[\triangledown\mathbf{u}]-\gamma\mathbf{F}^{-T}-\mu\triangledown\mathbf{u}\right\}=\mathbf{0},
\end{equation}
\begin{equation}\label{3.28}
\triangledown{\mathbf{u}}\cdot \mathbf{F}^{-T}=0,
\end{equation}
\begin{equation}\label{3.29}
u_\alpha=0 \quad \textrm{at} \  X_\alpha=\pm l_\alpha, \quad \alpha=1,2,
\end{equation}
\begingroup
    \fontsize{11.2pt}{12pt}\selectfont
\begin{equation}\label{3.30}
\left\{p\mathbf{F}^{-T}\triangledown\mathbf{u}^T\mathbf{F}^{-T}+\frac{\partial^2W}{\partial\mathbf{F}^2}[\triangledown\mathbf{u}]-\gamma\mathbf{F}^{-T}-\mu\triangledown\mathbf{u}\right\}\mathbf{e}_{\alpha}=\bar{z}_{\alpha}\mathbf{e}_{\alpha} \quad \textrm{at}\  X_\alpha=\pm l_\alpha, \ \alpha=1,2\ (\textrm{no sum}),
\end{equation}
\endgroup
\begin{equation}\label{3.31}
\left\{p\mathbf{F}^{-T}\triangledown\mathbf{u}^T\mathbf{F}^{-T}+\frac{\partial^2W}{\partial\mathbf{F}^2}[\triangledown\mathbf{u}]-\gamma\mathbf{F}^{-T}-\mu\triangledown\mathbf{u}\right\}\mathbf{e}_{3}=\mathbf{0} \quad \textrm{at}\  X_3=0,
\end{equation}
\begin{equation}\label{3.32}
\mathbf{u(X)=0}\quad \textrm{at}\  X_3=-l_3, \ \textrm{and}
\end{equation}
\begin{equation}\label{3.33}
\int_{\mathcal{B}}\left|\triangledown\mathbf{u}\right|^2dV=1,
\end{equation}
where $\gamma$ and $\mu$ are the Lagrange multipliers with respect to the constraint of incompressibility $\eqref{RuF}$ and the normalization condition $\eqref{Rnormal}$, respectively, and $\bar{z}_{\alpha}$ is the increment of the normal stress $z_{\alpha}$ on the side surface. To ensure the completeness of the boundary-value problem, equations $\eqref{RuF}$-$\eqref{Ru_B_C_2}$ and $\eqref{Rnormal}$ are rewritten here. Note that the constant Lagrange multiplier $\mu$ is connected to the value of the integral on the left-hand side of $\eqref{R>mu}$ for the solution ${\bf u}$ of this boundary-value problem \eqref{3.27}-\eqref{3.33}. Indeed, the solution ${\bf u}$ of the boundary-value problem \eqref{3.27}-\eqref{3.33} is the particular ${\bf u}$ making the equality hold in \eqref{R>mu}. And $\mu$ is equal to the value of the integral on the left-hand side of $\eqref{R>mu}$ for this particular ${\bf u}$. \footnote{Taking the inner produce of $\mathbf{u}$ and $\eqref{3.27}$, integrating the resulting equation on the domain $\mathcal{B}$, and using $\eqref{3.28}$-$\eqref{3.33}$, we find that the value of the integral on the left-hand side of $\eqref{R>mu}$ is exactly $\mu$ at a solution of the boundary-value problem $\eqref{3.27}$-$\eqref{3.33}$. }

We thus conclude that the quadratic integral inequality $\eqref{3.24}$ holds if and only if $\mu\ge0$ for all solutions of the boundary-value problem $\eqref{3.27}$-$\eqref{3.33}$. Based on the above statement, we have a direct criterion of the surface instability of an elastic block subject to biaxial loads:

{\it Consider a deformation that is the solution to the boundary-value problem consisting of equations \eqref{RE_L_Equation}-\eqref{RNatural_B_C_2} along with the constraint $\eqref{RD_G}$ and the kinematical boundary conditions $\eqref{RB_C_1}$-$\eqref{RB_C_2}$. 
If the eigenvalue $\mu$ is always nonnegative, $\mu \ge 0$, for all the solutions of the eigenvalue problem $\eqref{3.27}$-$\eqref{3.33}$ at that deformation, we claim that the deformation of the elastic block is stable. Otherwise, the deformation is unstable and the one has the lowest negative eigenvalue is more likely to make the deformation unstable. }

Note that the above stability criterion is valid for surface instability of a deformed block of any elastic materials. Also, the deformation of the deformed block is only determined by the boundary-value problem, which can be either homogeneous or inhomogeneous.
\section{Surface instability of a homogeneously deformed neo-Hookean block} \label{section4}
We now consider surface instability of homogeneous deformation of a neo-Hookean material as an example to carry out our stability analysis. The strain-energy function of neo-Hookean solids is 
\begin{equation}\label{RW_n_H}
W(\mathbf{F})=\frac{c}{2}(|\mathbf{F}|^2-3),
\end{equation}
where constant $c$ is the shear modulus of infinitesimal deformation. Then,
\begin{equation}\label{RP_W_n_H}
\frac{\partial W}{\partial\mathbf{F}}=c\mathbf{F},\,\frac{\partial^2 W}{\partial\mathbf{F}^2}=c\mathbf{I}_4,
\end{equation}
where $\mathbf{I}_4$ is the identity fourth order tensor. 

\subsection{Homogeneous deformation}

The equilibrium equation $\eqref{RE_L_Equation}$ and the boundary conditions $\eqref{RNatural_B_C}$ and $\eqref{RNatural_B_C_2}$ become
\begin{equation}\label{RE_L_n_H}
\textrm{Div}\left(c\mathbf{F}-p\mathbf{F}^{-T}\right)=\mathbf{0},
\end{equation}
\begin{equation}\label{RN_B_C_n_H}
\left(c\mathbf{F}-p\mathbf{F}^{-T}\right)\mathbf{e}_{\alpha}=z_{\alpha}\mathbf{e}_{\alpha}\quad \textrm{at} \  X_\alpha=\pm l_\alpha, \quad \alpha=1,2\ (\textrm{no sum}),\ \textrm{and}
\end{equation}
\begin{equation}\label{RN_B_C_n_H_2}
\left(c\mathbf{F}-p\mathbf{F}^{-T}\right){\mathbf{e}_{3}}=\mathbf{0} \quad \textrm{at}\ X_3=0.
\end{equation}
\indent A trivial solution, which corresponds to a homogeneous deformation, to equations $\eqref{RE_L_n_H}$-$\eqref{RN_B_C_n_H_2}$, the constraint $\eqref{RD_G}$ and the displacement boundary conditions $\eqref{RB_C_1}$-$\eqref{RB_C_2}$, is given by
\begin{equation}\label{Rx_H_S}
\mathbf{x}_{0}(\mathbf{X})=\lambda_1X_1\mathbf{e}_1+\lambda_2X_2\mathbf{e}_2+\frac{1}{\lambda_1\lambda_2}X_3\mathbf{e}_3,
\end{equation}
\begin{equation}\label{Rp_H_S}
p_0 (\mathbf{X})=\frac{c}{\lambda_1^2\lambda_2^2},
\end{equation}
and
\begin{equation}
z_{\alpha} = c \lambda_{\alpha} - \frac{c}{\lambda_{\alpha}} \frac{1}{\lambda_1^2 \lambda_2^2}, \quad \alpha=1,2\ (\textrm{no sum}).
\end{equation}

The corresponding deformation gradient of the homogeneous deformation ${\bf x}_0$ in \eqref{Rx_H_S} is
\begin{equation}\label{RD_G_H_S}
\mathbf{F}_0 (\mathbf{X}) = \nabla {\bf x}_0= \lambda_1\mathbf{e}_1\otimes\mathbf{e}_1+\lambda_2\mathbf{e}_2\otimes\mathbf{e}_2+\frac{1}{\lambda_1\lambda_2}\mathbf{e}_3\otimes\mathbf{e}_3.
\end{equation}

\subsection{Solution of the eigenvalue problem associated with the stability}\label{sec3.3}
\indent To examine the stability of the homogeneous solution \eqref{Rx_H_S}-\eqref{RD_G_H_S}, we now have to solve the eigenvalue problem associated with the stability. Substituting the homogeneous solution \eqref{Rx_H_S}-\eqref{RD_G_H_S} into the boundary-value problem $\eqref{3.27}$-$\eqref{3.33}$, we have
\begin{equation}\label{3.42}
\textrm{Div}\left(c\triangledown\mathbf{u}-\gamma\mathbf{F}_0^{-T}-\mu\triangledown\mathbf{u}\right)=\mathbf{0},
\end{equation}
\begin{equation}\label{3.43}
\triangledown{\mathbf{u}}\cdot \mathbf{F}_0^{-T}=0,
\end{equation}
\begin{equation}\label{3.44}
u_\alpha=0 \quad \textrm{at} \  X_\alpha=\pm l_\alpha, \quad \alpha=1,2,
\end{equation}
\begingroup
    \fontsize{11.3pt}{12pt}\selectfont
\begin{equation}\label{3.45}
\left(p_0 \mathbf{F}_0^{-T}\triangledown\mathbf{u}^T\mathbf{F}_0^{-T}+c\triangledown\mathbf{u}-\gamma\mathbf{F}_0^{-T}-\mu\triangledown\mathbf{u}\right)\mathbf{e}_{\alpha}=\bar{z}_{\alpha}\mathbf{e}_{\alpha} \quad \textrm{at} \  X_\alpha=\pm l_\alpha, \ \alpha=1,2 \ (\textrm{no sum}),
\end{equation}
\endgroup
\begin{equation}\label{3.46}
\left(p_0 \mathbf{F}_0^{-T}\triangledown\mathbf{u}^T\mathbf{F}_0^{-T}+c\triangledown\mathbf{u}-\gamma\mathbf{F}_0^{-T}-\mu\triangledown\mathbf{u}\right)\mathbf{e}_{3}=\mathbf{0} \quad \textrm{at}\  X_3=0,\ \textrm{and}
\end{equation}
\begin{equation}\label{3.47}
\mathbf{u(X)=0}  \quad \textrm{at}\  X_3=-l_3,
\end{equation}
along with $\eqref{3.33}$.

We now seek solutions, consisting of $\mathbf{u}$ and $\gamma$, of the differential equations $\eqref{3.42}$ and $\eqref{3.43}$ that satisfy the boundary conditions $\eqref{3.44}$ and $\eqref{3.45}$. Such solutions admit Fourier expansions of the form
\begingroup
    \fontsize{11.6pt}{14pt}\selectfont
\begin{align}\label{RF_u}
\mathbf{u(X)}&=\sum_{m,n=0}^{\infty}[\sin(mk_1X_1)\cos(nk_2X_2)U_{mn}(X_3)\mathbf{e}_1+\cos(mk_1X_1)\sin(nk_2X_2)V_{mn}(X_3)\mathbf{e}_2\nonumber\\
&\quad\quad\quad\quad+\cos(mk_1X_1)\cos(nk_2X_2)W_{mn}(X_3)\mathbf{e}_3],\\
\label{RF_gamma}
\gamma(\mathbf{X})&=\sum_{m,n=0}^{\infty}\cos(mk_1X_1)\cos(nk_2X_2)R_{mn}(X_3),
\end{align}
\endgroup
where
\begin{equation}\label{Rk}
k_{\alpha}=\frac{\pi}{l_{\alpha}},\quad \alpha=1,2,
\end{equation}
and, from the full Fourier expansion, we have retained only those terms satisfying the boundary conditions $\eqref{3.44}$ and $\eqref{3.45}$.\\
Substituting $\eqref{RF_u}$ and $\eqref{RF_gamma}$ into $\eqref{3.42}$ and $\eqref{3.43}$ and introducing
\begin{equation}\label{Rkmn}
k_{mn}=\sqrt{m^2k_1^2+n^2k_2^2},
\end{equation}
we obtain the following differential equations
\begin{equation}
\left.
\begin{aligned}\label{3.52}
&(c-\mu)\frac{\mathrm d^2U_{mn}(X_3)}{\mathrm dX_3^2}=(c-\mu)k_{mn}^2U_{mn}(X_3)-\frac{mk_1}{\lambda_1}R_{mn}(X_3),\\
&(c-\mu)\frac{\mathrm d^2V_{mn}(X_3)}{\mathrm dX_3^2}=(c-\mu)k_{mn}^2V_{mn}(X_3)-\frac{nk_2}{\lambda_2}R_{mn}(X_3),\\
&(c-\mu)\frac{\mathrm d^2W_{mn}(X_3)}{\mathrm dX_3^2}=(c-\mu)k_{mn}^2W_{mn}(X_3)+\lambda_1\lambda_2\frac{\mathrm dR_{mn}(X_3)}{\mathrm dX_3},\\
\end{aligned}
\right\}
\end{equation}
and
\begin{equation}\label{3.53}
\lambda_1\lambda_2\frac{\mathrm dW_{mn}(X_3)}{\mathrm dX_3}+\frac{mk_1}{\lambda_1}U_{mn}(X_3)+\frac{nk_2}{\lambda_2}V_{mn}(X_3)=0.
\end{equation}
Let us define the following forms
\begin{equation}\label{Ru_hat}
\begin{pmatrix}
U_{mn}(X_3)\\
V_{mn}(X_3)\\
W_{mn}(X_3)
\end{pmatrix}
=\mathbf{E}_{mn}e^{s_{mn}X_3}, \quad R_{mn}(X_3)=F_{mn}e^{s_{mn}X_3}.
\end{equation}
Here $\mathbf{E}_{mn}$ is a vector, $s_{mn}$ and $F_{mn}$ are scalars. All of them are independent of $X_3$.

Substituting $\eqref{Ru_hat}$ into $\eqref{3.52}$ and $\eqref{3.53}$, we have
\begin{equation}\label{3.55}
(c-\mu)(s_{mn}^2-k_{mn}^2)\mathbf{E}_{mn}+\lambda_1\lambda_2 F_{mn}(\boldsymbol{\eta}_{mn}-s_{mn}\mathbf{e}_3)=\mathbf{0}
\end{equation}
and
\begin{equation}\label{3.56}
\mathbf{E}_{mn}\cdot(\boldsymbol{\eta}_{mn}+s_{mn}\mathbf{e}_3)=0,
\end{equation}
where
\begin{equation}\label{Reta}
\boldsymbol{\eta}_{mn}=\frac{1}{\lambda_1\lambda_2}\left(\frac{mk_1}{\lambda_1}\mathbf{e}_1+\frac{nk_2}{\lambda_2}\mathbf{e}_2\right).
\end{equation}
\indent The case $\mu=c$ is non-consequential to the equation at hand since $c$ is assumed to be positive. When $\mu\neq c$, the system of algebraic equations $\eqref{3.55}$ and $\eqref{3.56}$ has 6 independent solutions:
\begin{align}%
s_{mn}^{(1)}&=k_{mn},\,\mathbf{E}_{mn}^{(1)}=\boldsymbol{\zeta}_{mn},\,F_{mn}^{(1)}=0;\nonumber\\
s_{mn}^{(2)}&=k_{mn},\,\mathbf{E}_{mn}^{(2)}=k_{mn}\boldsymbol{\eta}_{mn}-\eta_{mn}^2\mathbf{e}_3,\,F_{mn}^{(2)}=0;\nonumber\\
s_{mn}^{(3)}&=\eta_{mn},\,\mathbf{E}_{mn}^{(3)}=\lambda_1\lambda_2(\eta_{mn}\mathbf{e}_3-\boldsymbol{\eta}_{mn}),\,F_{mn}^{(3)}=(c-\mu)(\eta_{mn}^2-k_{mn}^2);\nonumber\\
s_{mn}^{(4)}&=-k_{mn},\,\mathbf{E}_{mn}^{(4)}=\boldsymbol{\zeta}_{mn},\,F_{mn}^{(4)}=0;\nonumber\\
s_{mn}^{(5)}&=-k_{mn},\,\mathbf{E}_{mn}^{(5)}=k_{mn}\boldsymbol{\eta}_{mn}+\eta_{mn}^2\mathbf{e}_3,\,F_{mn}^{(5)}=0;\nonumber\\
s_{mn}^{(6)}&=-\eta_{mn},\,\mathbf{E}_{mn}^{(6)}=-\lambda_1\lambda_2(\eta_{mn}\mathbf{e}_3+\boldsymbol{\eta}_{mn}),\,F_{mn}^{(6)}=(c-\mu)(\eta_{mn}^2-k_{mn}^2);\nonumber
\end{align}
where
\begin{equation}\label{Rzeta}
\boldsymbol{\zeta}_{mn}=\frac{nk_2}{\lambda_2}\mathbf{e}_1-\frac{mk_1}{\lambda_1}\mathbf{e}_2,\quad \eta_{mn}=|\boldsymbol{\eta}_{mn}|.
\end{equation}
\indent The general solution of the system of ordinary differential equations $\eqref{3.52}$ and $\eqref{3.53}$ is then given by
\begingroup
    \fontsize{11pt}{12pt}\selectfont
\begin{align}\label{RUmn}
\begin{pmatrix}
U_{mn}(X_3)\\
V_{mn}(X_3)\\
W_{mn}(X_3)
\end{pmatrix}&=\begin{aligned}
&[c_{mn}^{(1)}\boldsymbol{\zeta}_{mn}+c_{mn}^{(2)}(k_{mn}\boldsymbol{\eta}_{mn}-\eta_{mn}^2\mathbf{e}_3)]e^{k_{mn} X_3}+c_{mn}^{(3)}\lambda_1\lambda_2(\eta_{mn}\mathbf{e}_3-\boldsymbol{\eta}_{mn})e^{\eta_{mn} X_3}\\
&+[c_{mn}^{(4)}\boldsymbol{\zeta}_{mn}+c_{mn}^{(5)}(k_{mn}\boldsymbol{\eta}_{mn}+\eta_{mn}^2\mathbf{e}_3)]e^{-k_{mn} X_3}-c_{mn}^{(6)}\lambda_1\lambda_2(\eta_{mn}\mathbf{e}_3+\boldsymbol{\eta}_{mn})e^{-\eta_{mn} X_3},
\end{aligned}\\
\label{RRmn}
R_{mn}(X_3)&=(c-\mu)(\eta_{mn}^2-k_{mn}^2)(c_{mn}^{(3)}e^{\eta_{mn} X_3}+c_{mn}^{(6)}e^{-\eta_{mn} X_3}),
\end{align}
\endgroup
where $c_{mn}^{(i)},\,i=1,...,6$, are arbitrary constants.\\
Substituting $\eqref{RF_u}$ and $\eqref{RF_gamma}$ into the boundary conditions $\eqref{3.46}$ and $\eqref{3.47}$, we have
\begin{equation}
\left.
\begin{aligned}\label{3.61}
&(1-\frac{\mu}{c})\frac{\mathrm dU_{mn}(X_3)}{\mathrm dX_3}=\frac{mk_1}{\lambda_1^2 \lambda_2}W_{mn}(X_3),\\
&(1-\frac{\mu}{c})\frac{\mathrm dV_{mn}(X_3)}{\mathrm dX_3}=\frac{nk_2}{\lambda_1 \lambda_2^2}W_{mn}(X_3),\\
&(2-\frac{\mu}{c})\frac{\mathrm dW_{mn}(X_3)}{\mathrm dX_3}=\frac{\lambda_1\lambda_2}{c} R_{mn}(X_3),\\
\end{aligned}
\right\}\quad \textrm{at}\ X_3=0
\end{equation}
and
\begin{equation}\label{3.62}
U_{mn}(-l_3)=V_{mn}(-l_3)=W_{mn}(-l_3)=0.
\end{equation}
Substituting $\eqref{RUmn}$ and $\eqref{RRmn}$ into the boundary conditions $\eqref{3.61}$ and $\eqref{3.62}$ gives 6 equations, which can be written into a matrix form
\begin{equation}\label{RDjici}
\sum_{j,i=1}^{6}\mathsf{D}_{mn}^{(ji)}c_{mn}^{(i)}=0.
\end{equation}
Here the 6 by 6 coefficient matrix $\mathsf{D}_{mn}^{(ji)}$ in \eqref{RDjici} can be partitioned into four 3 by 3 blocks, such that
\begin{equation} \label{RDji}
\mathsf{D}_{mn}^{(ji)} =
\begin{pmatrix}
\mathsf{D}_{mn}^{a} & \mathsf{D}_{mn}^{b} \\
\mathsf{D}_{mn}^{c} & \mathsf{D}_{mn}^{d} 
\end{pmatrix},
\end{equation}
where
\begin{subequations}
\begin{align}
\label{RDji_1}
\mathsf{D}_{mn}^{a} & =
\begin{pmatrix}
(1-\frac{\mu}{c})k_{mn}\frac{nk_2}{\lambda_2} & {\upsilon_{mn}}\frac{mk_1}{\lambda_1^2 \lambda_2} & -(2-\frac{\mu}{c})\eta_{mn}\frac{mk_1}{\lambda_1} \\
-(1-\frac{\mu}{c})k_{mn}\frac{mk_1}{\lambda_1} & {\upsilon_{mn}}\frac{nk_2}{\lambda_1 \lambda_2^2} & -(2-\frac{\mu}{c})\eta_{mn}\frac{nk_2}{\lambda_2} \\
0 & -(2-\frac{\mu}{c})k_{mn}\eta_{mn}^2 & \lambda_1\lambda_2 {\upsilon_{mn}}
\end{pmatrix}, \\
\label{RDji_2}
\mathsf{D}_{mn}^{b} &=
\begin{pmatrix}
-(1-\frac{\mu}{c}) k_{mn}\frac{n k_2}{\lambda_2} & -{\upsilon_{mn}}\frac{mk_1}{\lambda_1^2 \lambda_2} & (2-\frac{\mu}{c})\eta_{mn}\frac{mk_1}{\lambda_1} \\
(1-\frac{\mu}{c})k_{mn}\frac{mk_1}{\lambda_1} & -{\upsilon_{mn}}\frac{nk_2}{\lambda_1 \lambda_2^2} & (2-\frac{\mu}{c})\eta_{mn}\frac{nk_2}{\lambda_2} \\
 0  & -(2-\frac{\mu}{c})k_{mn}\eta_{mn}^2 & \lambda_1\lambda_2 {\upsilon_{mn}}
\end{pmatrix}, \\
\label{RDji_3}
\mathsf{D}_{mn}^{c} & =
\begin{pmatrix}
\frac{nk_2}{\lambda_2}e^{-k_{mn}l_3} & k_{mn}\frac{mk_1}{\lambda_1^2 \lambda_2}e^{-k_{mn}l_3} & -\frac{mk_1}{\lambda_1}e^{-\eta_{mn}l_3} \\
-\frac{mk_1}{\lambda_1}e^{-k_{mn}l_3} & k_{mn}\frac{nk_2}{\lambda_1 \lambda_2^2}e^{-k_{mn}l_3} & -\frac{nk_2}{\lambda_2}e^{-\eta_{mn}l_3} \\
0 & -\eta_{mn}^2 e^{-k_{mn}l_3} & \lambda_1 \lambda_2 \eta_{mn} e^{-\eta_{mn} l_3}
\end{pmatrix}, \\
\label{RDji_4}
\mathsf{D}_{mn}^{d} & =
\begin{pmatrix}
\frac{nk_2}{\lambda_2}e^{k_{mn}l_3} &  k_{mn}\frac{mk_1}{\lambda_1^2 \lambda_2}e^{k_{mn}l_3} & -\frac{mk_1}{\lambda_1}e^{\eta_{mn}l_3} \\
 -\frac{mk_1}{\lambda_1}e^{k_{mn}l_3} & k_{mn}\frac{nk_2}{\lambda_1 \lambda_2^2}e^{k_{mn}l_3} & -\frac{n k_2}{\lambda_2}e^{\eta_{mn}l_3} \\
0 & \eta_{mn}^2 e^{k_{mn}l_3} & -\lambda_1 \lambda_2 \eta_{mn} e^{\eta_{mn} l_3}
\end{pmatrix},
\end{align}
\end{subequations}
where ${\upsilon_{mn}}=(1-\frac{\mu}{c})k_{mn}^2+\eta_{mn}^2$.\\
\indent The necessary condition for non-zero solutions of $c_{mn}^{(i)}$, $i=1,2,...,6,$ is that the determinant of the matrix $\mathsf{D}_{mn}^{(ji)}$ in \eqref{RDji} must vanish. In terms of the dimensionless wavenumber
\begin {equation}\label {xi_bar}
\bar{k}_{mn}=k_{mn} l_3,
\end {equation}
the characteristic equation from $\det {\mathsf{D}_{mn}^{(ji)}} = 0$ in \eqref{RDji} is obtained as
\begin{align}\label {C_E_4}
&4(\frac{\mu}{c}-1){\lambda_1^4 \lambda_2^4}{k_{mn}^6}{\eta_{mn}^{5}}\cosh{\bar{k}_{mn}}\times\nonumber\\
&\left\{({t_{mn}}+1)\left[(2-\frac{\mu}{c})^2 {t_{mn}^3}+{({t_{mn}^2}+1-\frac{\mu}{c})}^2\right]\cosh [(1-{t_{mn}})\bar{k}_{mn}]-4{t_{mn}^2}(2-\frac{\mu}{c}){({t_{mn}^2}+1-\frac{\mu}{c})}\right.\nonumber\\
&\quad\left.+({t_{mn}}-1)\left[(2-\frac{\mu}{c})^2 t_{mn}^3-{(t_{mn}^2+1-\frac{\mu}{c})}^2\right]\cosh [(1+{t_{mn}})\bar{k}_{mn}]\right\}=0,
\end{align}
with
\begin{equation}\label{t}
{t_{mn}}=\frac{\eta_{mn}}{k_{mn}}>0,
\end{equation}
where $k_{mn}$ is defined by $\eqref{Rkmn}$ and $\eta_{mn}$ is defined by $\eqref{Rzeta}$.
In the following, we will find the stability condition by discussing the characteristic equation $\eqref{C_E_4}$.

\indent Recall that whether the second variation condition $\eqref{R>mu}$ holds is equivalent to whether all the eigenvalues are non-negative. It is obvious that one root of $\mu$ in $\eqref{C_E_4}$ is $c$ that is non-consequential due to the assumption of a positive $c$. Now the signs of the other two real roots of $\mu$ are interested and $\eqref{C_E_4}$ is reduced to
\begin{align}\label {C_E_4_1}
&({t_{mn}}+1)\left[(2-{\bar\mu})^2 {t_{mn}^3}+{({t_{mn}^2}+1-{\bar\mu})}^2\right]\cosh [(1-{t_{mn}})\bar{k}_{mn}]-4{t_{mn}^2}(2-{\bar\mu}){({t_{mn}^2}+1-{\bar\mu})}\nonumber\\
&\quad+({t_{mn}}-1)\left[(2-{\bar\mu})^2 t_{mn}^3-{(t_{mn}^2+1-{\bar\mu})}^2\right]\cosh [(1+{t_{mn}})\bar{k}_{mn}]=0,
\end{align}
where
\begin{equation}\label {bar_mu_c}
{\bar\mu}=\frac{\mu}{c}.
\end{equation} 
Note that $\eqref{C_E_4_1}$ is a quadratic equation of ${\bar\mu}$ at any given pair $(m,n)$, such that
\begin{equation}\label {mu_quad}
a_{mn}\bar\mu^2+b_{mn}\bar\mu+c_{mn}=0,
\end{equation} 
with
\begin{subequations} \label {abc_t_xi}
\begin{align}
\label {a_t_xi}
a_{mn}&=\hat a(t_{mn},\bar{k}_{mn}) \nonumber\\
           &=(t_{mn}+1)^2(t_{mn}^2-t_{mn}+1)\text{cosh}[(1-t_{mn})\bar{k}_{mn}]-4t_{mn}^2\nonumber\\
&\quad +(t_{mn}-1)^2(t_{mn}^2+t_{mn}+1)\text{cosh}[(1+t_{mn})\bar{k}_{mn}],\\
\label {b_t_xi}
b_{mn}&=\hat b(t_{mn},\bar{k}_{mn}) \nonumber\\
           &=-2 \left\{(t_{mn}+1)^2(2t_{mn}^2-t_{mn}+1)\text{cosh}[(1-t_{mn})\bar{k}_{mn}] -2t_{mn}^2(t_{mn}^2+3)\right.\nonumber\\
&\quad \left.+(t_{mn}-1)^2(2t_{mn}^2+t_{mn}+1)\text{cosh}[(1+t_{mn})\bar{k}_{mn}]\right\},\\
\label {c_t_xi}
c_{mn}& =\hat c(t_{mn},\bar{k}_{mn}) \nonumber\\
           &=(t_{mn}+1)^2(t_{mn}^3+3t_{mn}^2-t_{mn}+1)\text{cosh}[(1-t_{mn})\bar{k}_{mn}]-8t_{mn}^2(t_{mn}^2+1)\nonumber\\
           &\quad-(t_{mn}-1)^2(t_{mn}^3-3t_{mn}^2-t_{mn}-1)\text{cosh}[(1+t_{mn})\bar{k}_{mn}],
\end{align}
\end{subequations}
where $\bar{k}_{mn}$ and $t_{mn}$ are defined by $\eqref{xi_bar}$ and $\eqref{t}$, respectively. The discriminant of $\eqref{mu_quad}$ is
\begin{align}\label{discriminant0}
\triangle_{mn}&=\hat \triangle(t_{mn},\bar{k}_{mn})=[\hat b(t_{mn},\bar{k}_{mn})]^2 -4 [\hat a(t_{mn},\bar{k}_{mn})] [\hat c(t_{mn},\bar{k}_{mn})]\nonumber\\
                    &=4t_{mn}^3 (t_{mn}^2-1)^2 \nonumber\\
                    & \quad \times \left\{(t_{mn}-1)^2\text{cosh}^2[(1+t_{mn})\bar{k}_{mn}]-(t_{mn}+1)^2\text{cosh}^2[(1-t_{mn})\bar{k}_{mn}]+4t_{mn}\right\}.
\end{align} 
From the property of the hyperbolic cosine function, we have the following inequalities
\begin{subequations} \label {abc_t_xi_ine}
\begin{align}
\label {a_t_xi_ine}
\hat a(t_{mn},\bar{k}_{mn})&>\hat a(t_{mn},0)=2(t_{mn}^2-1)^2>0,\\
\label {b_t_xi_ine}
\hat b(t_{mn},\bar{k}_{mn})&<\hat b(t_{mn},0)=-4(t_{mn}^2-1)^2<0,\\
\label {discriminant}
\hat \triangle(t_{mn},\bar{k}_{mn})&>\hat \triangle(t_{mn},0)=0.
\end{align}
\end{subequations}

Hence, the quadratic equation $\eqref{mu_quad}$ has two distinct real roots and the quadratic formula is
\begin{equation}\label {two_roots}
\bar\mu_{1,2}=\frac{-b_{mn} \pm\sqrt{\triangle_{mn}}}{2a_{mn}}.
\end{equation} 

\section{Stability conditions and discussions} \label{section5}
Now the requirement of negative $\bar{\mu} = {\mu}/{c}$ in \eqref{two_roots}, the condition of surface instability leads to that the lower eigenvalue $\bar\mu_2$ should be negative
\begin{equation}\label{negative_mu2}
\bar\mu_2=\frac{-b_{mn} -\sqrt{\triangle_{mn}}}{2a_{mn}} < 0.
\end{equation}

Equation \eqref{negative_mu2} is the condition of surface instability of a homogeneously deformed block of neo-Hookean materials. It is clear from \eqref{abc_t_xi} and \eqref{discriminant0} that the value of $\bar\mu_2$ in \eqref{negative_mu2} directly depends on the values of $t_{mn}$ and ${\bar k}_{mn}$. In the following, we will show the variation of $\bar\mu_2$ with respect to $t_{mn}$ and ${\bar k}_{mn}$, and then the values of $\bar\mu_2$ in terms of the two stretches $\lambda_1$ and $\lambda_2$. With the condition of surface instability, we can finally get the stability and instability regions on the $\lambda_1 - \lambda_2$ plane.

\subsection{Value of $\bar\mu_2$ in terms of $t_{mn}$ and ${\bar k}_{mn}$}
With the definitions of $t_{mn} = \frac{\eta_{mn}}{k_{mn}} > 0 $ in \eqref{t} and ${\bar k}_{mn} =  k_{mn} l_3 > 0$ in \eqref{Rkmn}, we know that $t_{mn}$ and ${\bar k}_{mn}$ are actually discrete variables depending on the integers $m$ and $n$. 

To investigate the trend of the change of $\bar\mu_2$ in \eqref{negative_mu2}, we take continuous variables $t \in {\mathbb R}^+$ and $k \in {\mathbb R}^+$ rather than discrete variables $t_{mn}$ and ${\bar k}_{mn}$ in \eqref{abc_t_xi}-\eqref{negative_mu2}. Thus $\bar\mu_2$ in \eqref{negative_mu2} is a function of two continuous variables $t \in {\mathbb R}^+$ and $k \in {\mathbb R}^+$. Actually, $t_{mn}$ and ${\bar k}_{mn}$ can be approximately assumed to be continuous at a block with sufficiently large $l_i, i=1,2,3$.

\begin{figure}[h]
\centering
\subfigure[]{%
        \includegraphics[width=3.1in]{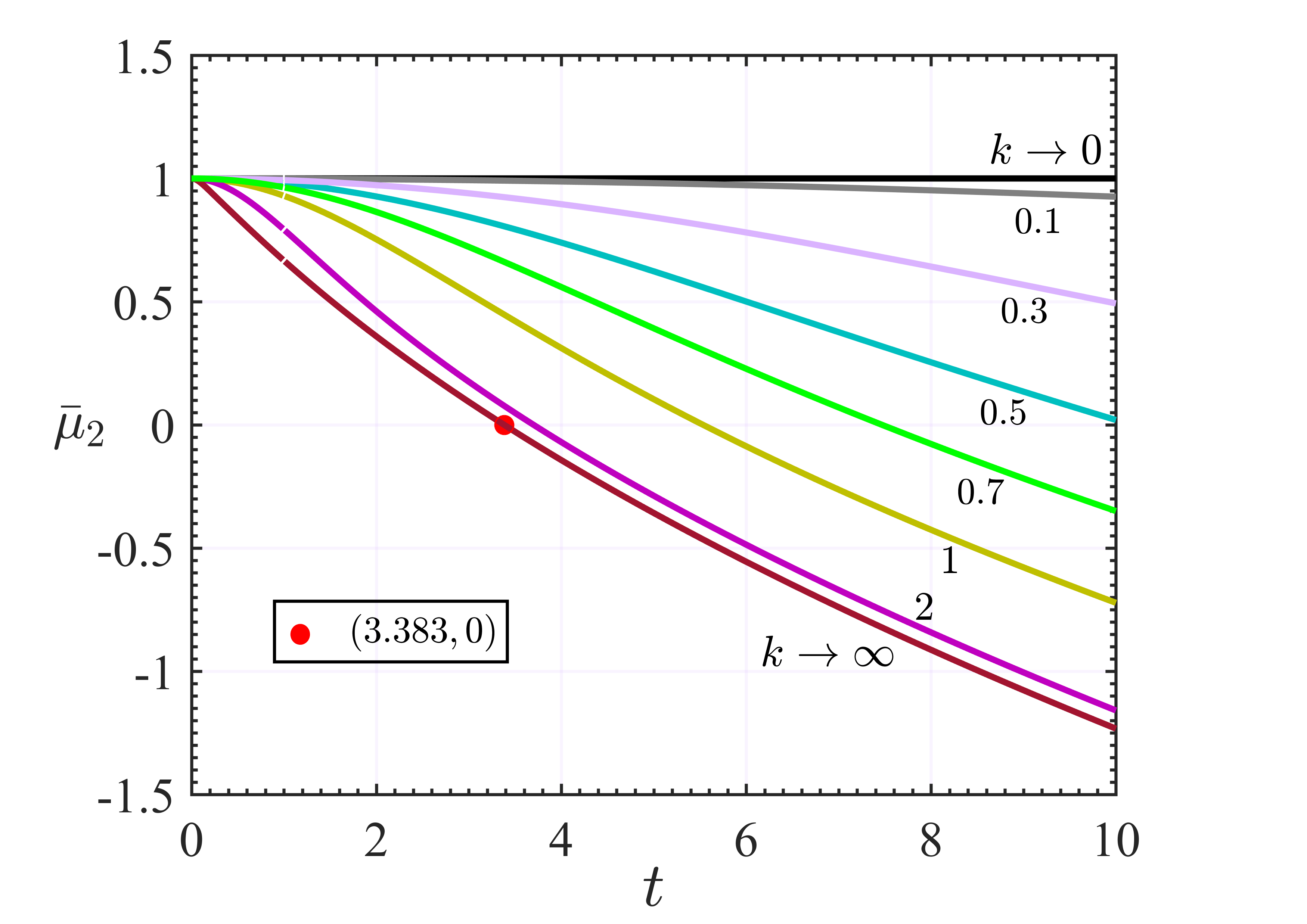}
       \label{fig2a} } 
\subfigure[]{%
        \includegraphics[width=3.1in]{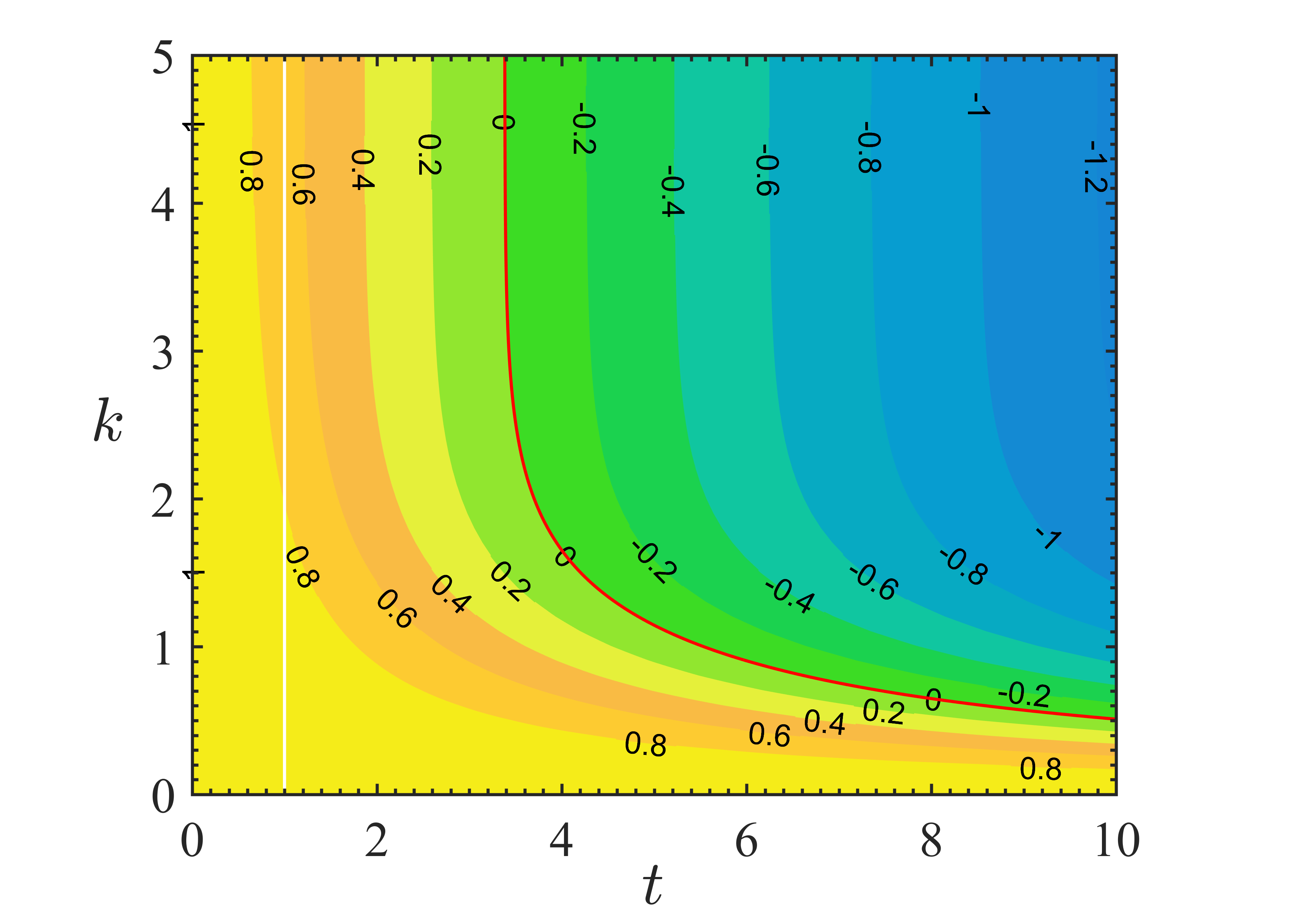}
       \label{fig2b} }
 \caption{Value of the eigenvalue $\bar\mu_2$ in \eqref{negative_mu2}: (a) $\bar\mu_2$ {\it vs}. $t$ under various $k$. For $k \to 0$, the eigenvalue ${\bar\mu}_2 \to 1$ in a sufficiently large range $(0,t)$. For $k \to \infty$, the eigenvalue ${\bar\mu}_2$ decreases from 1 to 0 as $t$ increases from 0 to $3.383$; (b) $\bar\mu_2$ in the $t - k$ plane. The straight line $t=1$ corresponds to $\eta_{mn}=k_{mn}$, which is an undeformed state $\lambda_1 =\lambda_2 =1$ whose stability can be deduced from that of the neighboring states. The solid red curve in (b) is the summation of all the points $(t,k)$ at which the eigenvalue $\bar \mu_2$ is zero.}
 \label{fig2}
 \end{figure}

In Figure \ref{fig2a}, we plot the variation of $\bar\mu_2$ in \eqref{negative_mu2} with a continuous $t$ under various $k$. For a prescribed $k$, the normalized eigenvalue $\bar\mu_2$ decreases monotonically from $1$ to negative with the increase of $t$ from 0 to a sufficiently large value. In addition, at any prescribed $t$, a larger $k$ corresponds to a smaller $\bar\mu_2$, indicating that $\bar\mu_2$ also decreases monotonically with $k$. 

Figure \ref{fig2b} shows the values of $\bar\mu_2$ in the $t-k$ plane. The eigenvalue $\bar\mu_2$ is positive on the bottom left while it is negative on the top right. There exists only one curve on which the eigenvalue $\bar\mu_2$ is zero in the $t-k$ plane, namely
\begin{equation}\label{zero-mu-curve}
\mathcal F (t, k) = 0, \quad 3.383 < t < \infty, 0 < k < \infty.
\end{equation}

The zero curve \eqref{zero-mu-curve} is represented by a solid red curve in Figure \ref{fig2b}. The eigenvalue $\bar\mu_2$ on the left-hand side of the zero curve is positive while $\bar\mu_2$ is negative on the right-hand side. 

From Figure \ref{fig2b} we know that \eqref{zero-mu-curve} is only curve of zero $\bar \mu_2$ and there is no branch bifurcating from this curve \eqref{zero-mu-curve}. With the implicit function theorem, there exists a function $\mathcal G: (3.383, \infty) \to \mathbb R^+$, such that
\begin{equation}\label{zero-mu-curve-k-t}
k=\mathcal G (t), \quad 3.383 < t < \infty,
\end{equation}
and
\begin{equation}\label{zero-mu-curve-k-t-1}
\mathcal F (t, \mathcal G (t)) = 0, \quad 3.383 < t < \infty.
\end{equation}

The limits of the function $\mathcal G$ are
\begin{equation}\label{zero-mu-curve-k-t-2}
\lim_{t \to 3.383^+} \mathcal G (t) = \infty \quad {\rm and} \quad \lim_{t \to \infty} \mathcal G (t) = 0.
\end{equation}

In the $t-k$ plane in Figure \ref{fig2b}, together with the function \eqref{zero-mu-curve-k-t}, the sign of the eigenvalue ${\bar\mu}_2$ is determined by
\begin{equation} \label{mu-p-0-n}
\left\{
\begin{aligned}
& {\bar\mu}_2 >0 \qquad {\rm if} \ (t, k) \in {\bf R}^+ = \{ (t, k) \in \mathbb R^+ \times \mathbb R^+: 0 < t < \infty, 0 < k < \mathcal G (t)\}, \\
& {\bar\mu}_2 =0 \qquad {\rm if} \ 0 < t < \infty \ {\rm and} \ k = \mathcal G (t), \\
& {\bar\mu}_2 <0 \qquad {\rm if} \ (t, k) \in {\bf R}^- = \{ (t, k) \in \mathbb R^+ \times \mathbb R^+: 0 < t < \infty, \mathcal G (t) < k < \infty\}.
\end{aligned}
\right.
\end{equation}
\subsection{Range of the ratio $t_{mn}$ for a given pair of stretches $(\lambda_1, \lambda_2)$}
At a give pair $(\lambda_1,\lambda_2)$, the ratio $t_{mn}$ in $\eqref{t}$ can be written explicitly as
\begin{equation}\label{t_mn_exp}
t_{mn}=\frac{1}{\lambda_1\lambda_2}\left[\frac{1}{\lambda_1^2}+\left(\frac{1}{\lambda_2^2}-\frac{1}{\lambda_1^2}\right)\frac{(n/l_2)^2}{(m/l_1)^2+(n/l_2)^2}\right]^{1/2}, \quad m, n \in \mathbb Z^{\ge}.
\end{equation}

\indent If $\lambda_1>\lambda_2$, the maximum $t_{mn}$ in $\eqref{t_mn_exp}$ is obtained at $m=0$ with any $n \in \mathbb Z^{+}$ while the minimum is achieved at $n=0$ with any $m \in \mathbb Z^{+}$. Similarly, if $\lambda_2>\lambda_1$, the maximum $t_{mn}$ is obtained at $n=0$ with any $m \in \mathbb Z^{+}$ while the minimum is achieved at $m=0$ with any $n \in \mathbb Z^{+}$. In particular, if $\lambda_2=\lambda_1$, $t_{mn}$ is independent of the wavenumbers. Thus, for a given pair of stretches $(\lambda_1, \lambda_2)$, $t_{mn}$ in $\eqref{t_mn_exp}$ must be in the range
\begin{equation}\label{t_mn_range}
\frac{\textrm{min}\{\lambda_1,\lambda_2\}}{\lambda_1^2\lambda_2^2}=t_a \le t_{mn} \le t_b= \frac{\textrm{max}\{\lambda_1,\lambda_2\}}{\lambda_1^2\lambda_2^2}.
\end{equation}

We define the set 
\begin{equation}\label{set-R-t-k}
{\bf R}_{\rm t} = \{ (t, k) \in \mathbb R^+ \times \mathbb R^+: t_a \le t \le t_b, 0 < k < \infty \}
\end{equation}
in which the values $(t,k)$ can be obtained for a given pair of stretches $(\lambda_1, \lambda_2)$.

Recall the range of $t \in (3.383, \infty)$ in the zero curve \eqref{zero-mu-curve}. Together with \eqref{t_mn_range} of the range of $t_{mn}$ for a given pair $(\lambda_1, \lambda_2)$, there are three cases:
\begin{subequations} \label{3cases-85}
\begin{align}
&{\text {Case I}}: \quad & [t_a, t_b]  & \subset (0, 3.383), & \\
&{\text {Case II}}: \quad & 3.383      & \in [t_a, t_b], & \\
&{\text {Case III}}: \quad & [t_a, t_b]  &\subset (3.383, \infty). & 
\end{align}
\end{subequations}

Combining Figure \ref{fig2} and the sign of ${\bar \mu}_2$ in \eqref{mu-p-0-n} as well as the three cases of different ranges of $[t_a, t_b]$ in \eqref{3cases-85}, we will illustrate the value of the eigenvalue ${\bar \mu}_2$ for a given pair $(\lambda_1, \lambda_2)$.
 
\subsection{Value of ${\bar \mu}_2$ in the range of the ratio $t_{mn}$}

Recall the defined sets ${\bf R}^+$ and ${\bf R}^-$ in \eqref{mu-p-0-n} and ${\bf R}_{\rm t}$ in \eqref{set-R-t-k}. In Case I, the set ${\bf R}_{\rm t}$ is a subset of ${\bf R}^+$ and then the intersection of sets ${\bf R}_{\rm t}$ and ${\bf R}^+$ is equal to ${\bf R}_{\rm t}$, namely
\begin{subequations}\label{case-I-set}
\begin{equation} \label{case-I-set-1}
\emptyset \neq {\bf R}_{\rm t} \subset {\bf R}^+, \quad {\bf R}_{\rm t}^+ = {\bf R}_{\rm t} \cap {\bf R}^+ = {\bf R}_{\rm t}.
\end{equation}
while the intersection of sets ${\bf R}_{\rm t}$ and ${\bf R}^-$ is an empty set, namely
\begin{equation} \label{case-I-set-2}
\emptyset = {\bf R}_{\rm t}^- = {\bf R}_{\rm t} \cap {\bf R}^-.
\end{equation}
\end{subequations}

\begin{figure}[H]
\centering
\subfigure[]{%
        \includegraphics[width=2.3in]{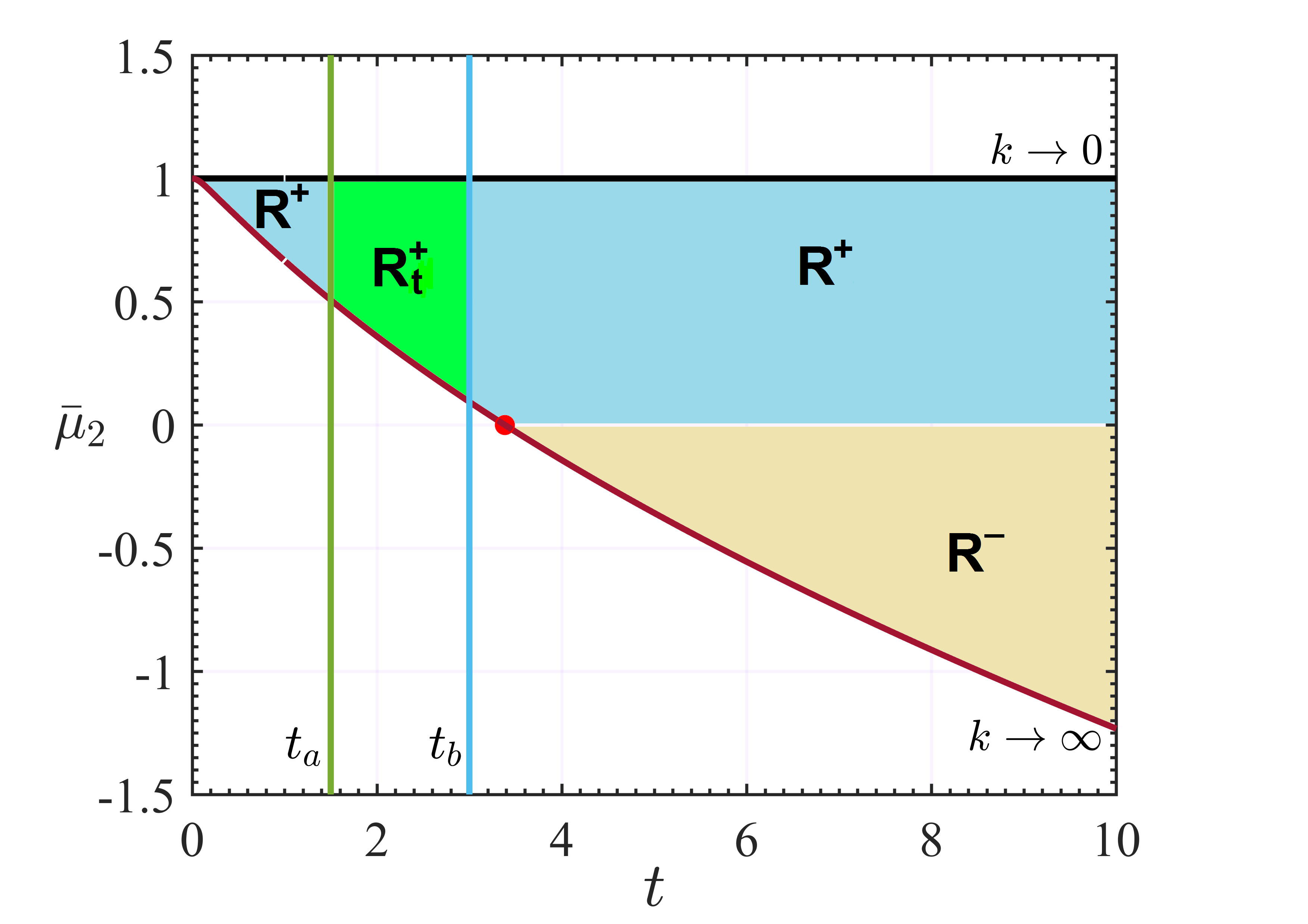}
       \label{fig3a} }
\subfigure[]{%
        \includegraphics[width=2.3in]{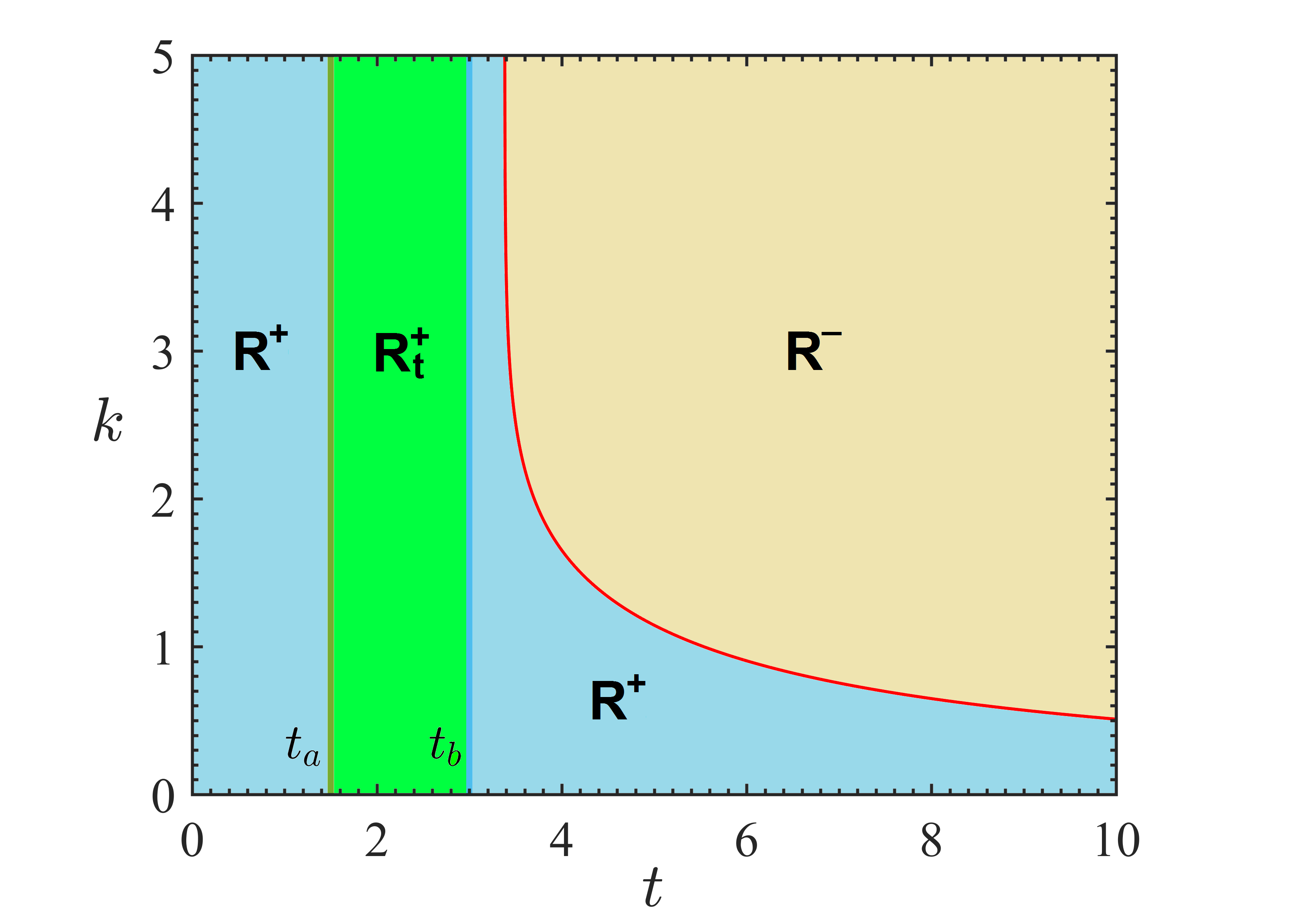}
       \label{fig3b} }
       \subfigure[]{%
        \includegraphics[width=2.3in]{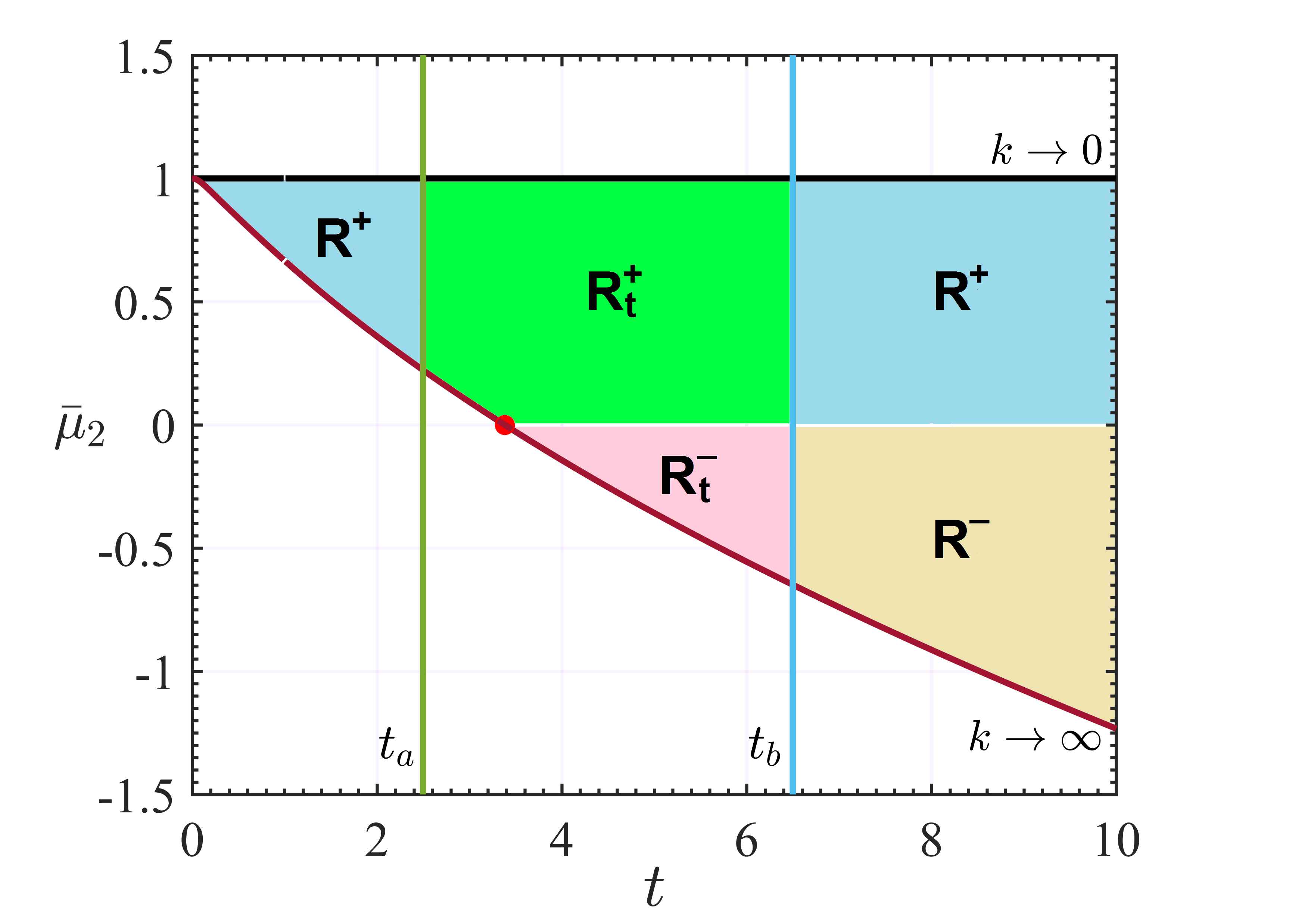}
       \label{fig3c} }
\subfigure[]{%
        \includegraphics[width=2.3in]{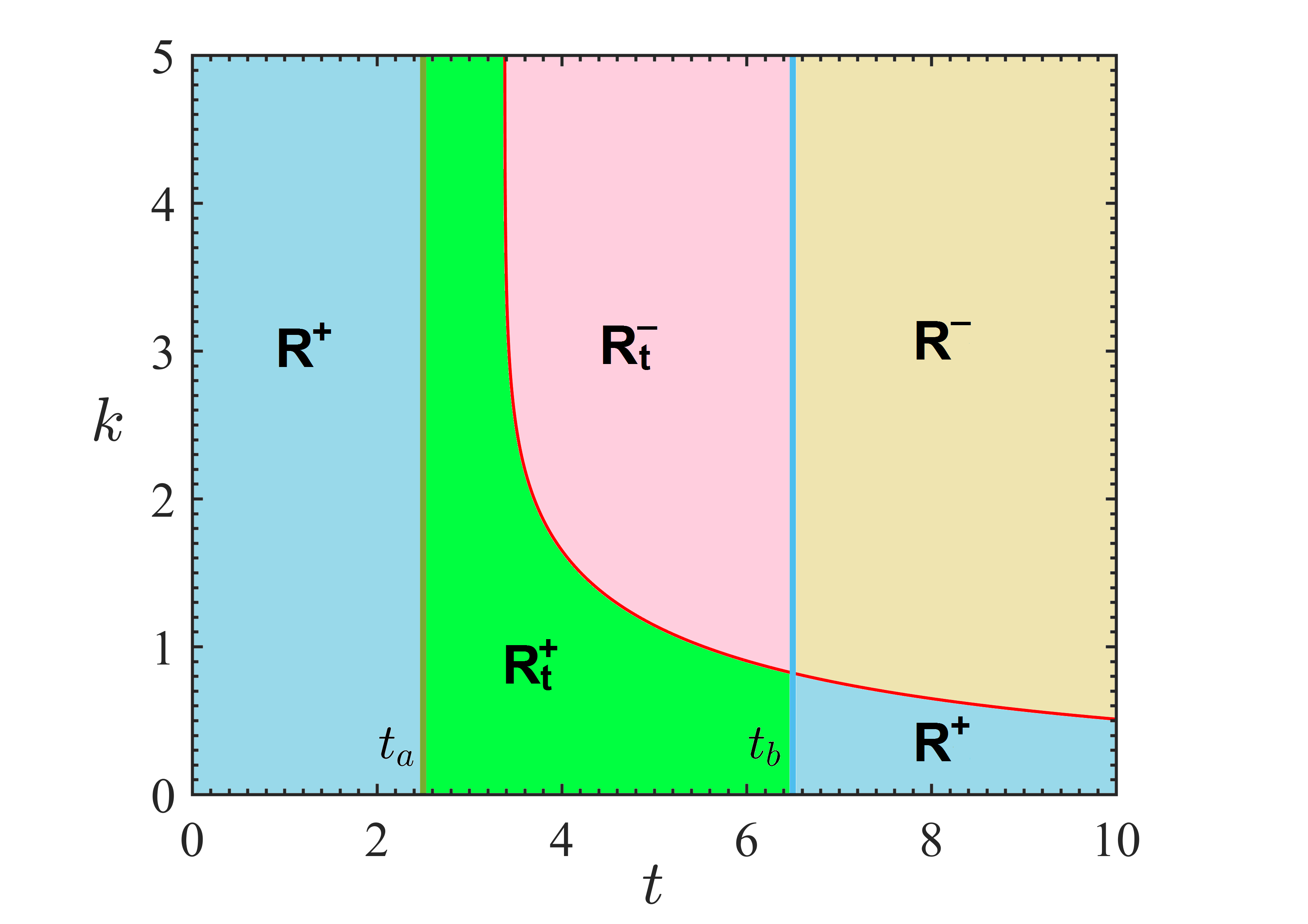}
       \label{fig3d} }
       \subfigure[]{%
        \includegraphics[width=2.3in]{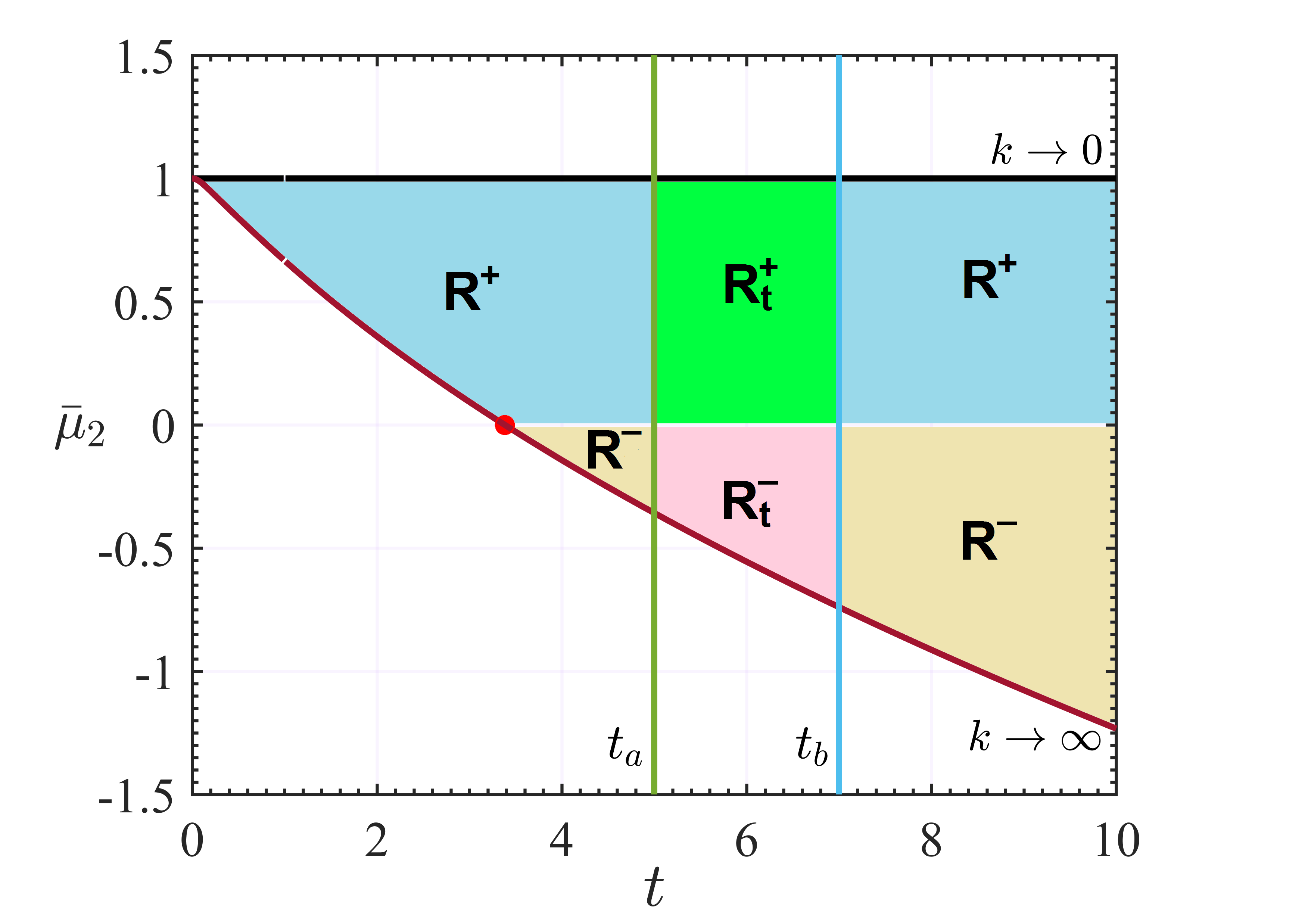}
       \label{fig3e} }
\subfigure[]{%
        \includegraphics[width=2.3in]{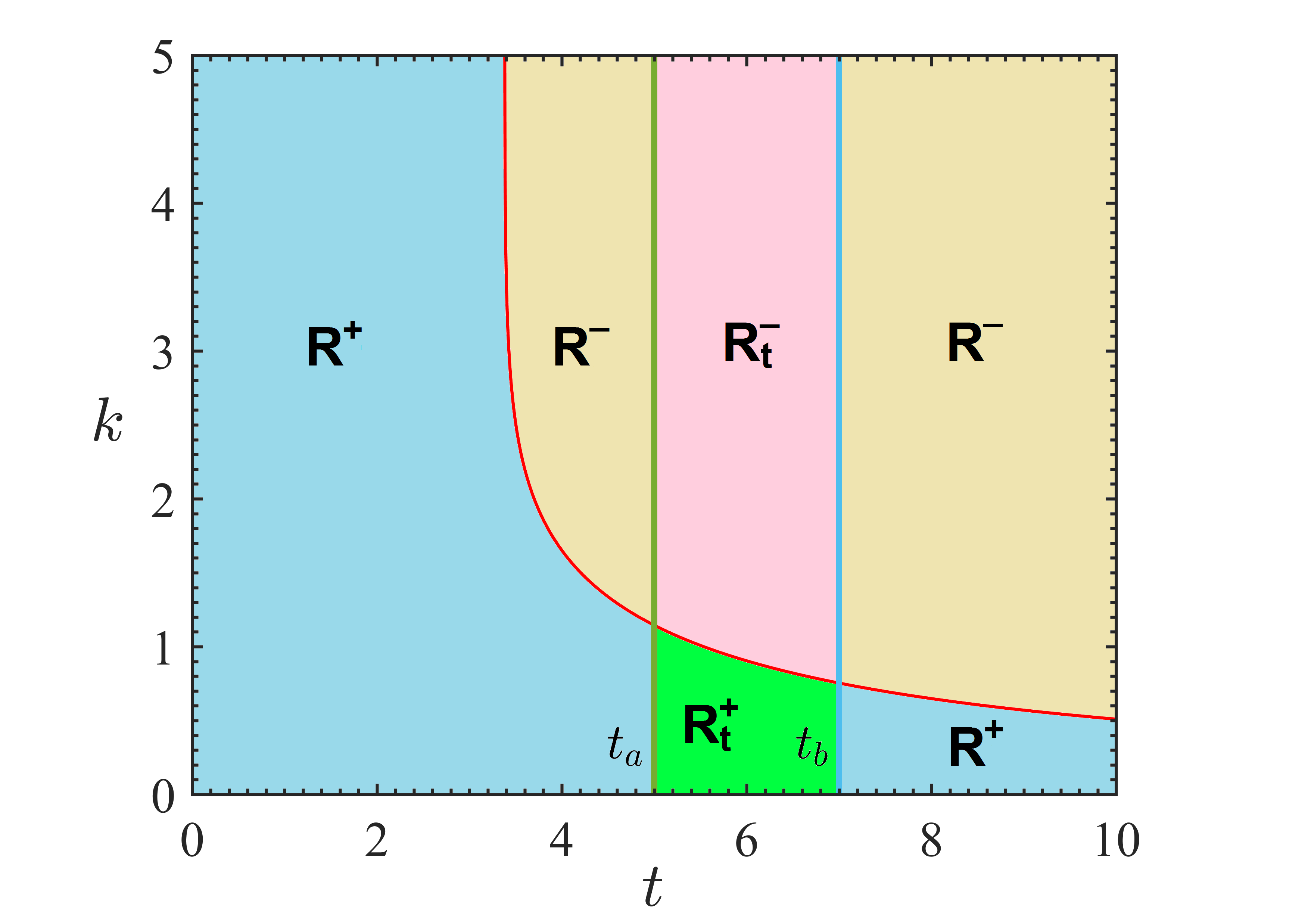}
       \label{fig3f} }
 \caption{Separations of the regions of three cases of the range of $t$ for a given pair $(\lambda_1, \lambda_2)$. (a) and (b) Case I: $[t_a, t_b]  \subset (0, 3.383)$;  (c) and (d) Case II: $3.383 \in [t_a, t_b]$; (e) and (f) Case III: $[t_a, t_b]  \subset (3.383, \infty)$. In these three cases, ${\bf R}^+$ denotes the region with positive eigenvalue while ${\bf R}^-$ denotes the region with negative eigenvalue. In particular, ${\bf R}_t^+$ and ${\bf R}_t^-$ denote, respectively, the regions with positive and negative eigenvalues for $t \in [t_a, t_b]$.} 
 \label{fig3}
 \end{figure}

The set operations, \eqref{case-I-set-1} and \eqref{case-I-set-2}, in Case I are also shown in Figures \ref{fig3a} and \ref{fig3b}.
Thus the set for negative eigenvalues ${\bar \mu}_2 (k,t)$ is empty in Case I (see \eqref{case-I-set-2} or Figure \ref{fig3b}), implying that the eigenvalue ${\bar \mu}_2 (k,t)$ is always positive for $(t,k) \in {\bf R}_{\rm t} \subset {\bf R}^+$. 

Alternatively, the minimum and maximum values of ${\bar \mu}_2$ for $(t, k) \in {\bf R}_{\rm t}$ in Case I are positive
\begin{equation}\label{case-1-extremum}
{\rm min} \ {\bar \mu}_2 (t, k) = {\bar \mu}_2 (t_b, \infty) > 0, \quad {\rm max} \ {\bar \mu}_2 (t, k) = {\bar \mu}_2 (t_b, 0) > 0,
\end{equation}
and
\begin{equation}\label{case-1-value-posi-0}
{\bar \mu}_2 (t, k) > 0, \quad {\rm for} \ (t, k) \in {\bf R}_{\rm t} \subset {\bf R}^+,
\end{equation}
which indicates that the homogeneously deformed elastic block is stable and there is no surface wrinkling. 

Again, recall the defined sets ${\bf R}^+$ and ${\bf R}^-$ in \eqref{mu-p-0-n} and ${\bf R}_{\rm t}$ in \eqref{set-R-t-k}. Both in Cases II and III, we have the intersections
\begin{equation} \label{case-II-III-set}
\emptyset \neq {\bf R}_{\rm t}^+ = {\bf R}_{\rm t} \cap {\bf R}^+, \quad \emptyset \neq {\bf R}_{\rm t}^- = {\bf R}_{\rm t} \cap {\bf R}^-.
\end{equation}

The non-empty set ${\bf R}_{\rm t}^- \neq \emptyset$ in \eqref{case-II-III-set} indicates that the negative eigenvalue ${\bar \mu}_2 (k,t)$ can be achieved for $(t,k) \in {\bf R}_{\rm t}^- \neq \emptyset$ in Cases II and III, see Figures \ref{fig3c} and \ref{fig3d} for Case II and Figures \ref{fig3e} and \ref{fig3f} for Case III. Alternatively, the minimum and maximum values of ${\bar \mu}_2$ for $(t, k) \in {\bf R}_{\rm t}$ in both Cases II and III have the properties
\begin{equation} \label{cases-2-3-extremum}
{\rm min} \ {\bar \mu}_2 (t, k) = {\bar \mu}_2 (t_b, \infty) < 0, \quad {\rm max} \ {\bar \mu}_2 (t, k) = {\bar \mu}_2 (t_b, 0) > 0.
\end{equation}

Since the negative eigenvalue indicates the negative second variation of the energy functional, the surface wrinkling will occur in Cases II and III rather in Case I.

\subsection{Stability and instability regions in the $\lambda_1 - \lambda_2$ plane}
\indent Combining the above three cases \eqref{3cases-85} and their extremum values \eqref{case-1-extremum} and \eqref{cases-2-3-extremum}, we have the stability and instability regions in terms of the stretches $\lambda_1$ and $\lambda_2$ as:
\begin{subequations}\label{sta--insta-condi-stretch}
\begin{align}
\label{sta--insta-condi-stretch-1}
& {\rm stable}, \, \qquad {\rm if} \ (\lambda_1,\lambda_2) \in \mathcal S_{\rm stable}=\left\{ (\lambda_1,\lambda_2) \in {\mathbb R}^+ \times {\mathbb R}^+: \frac{\textrm{max}\{\lambda_1,\lambda_2\}}{\lambda_1^2\lambda_2^2} < 3.383 \right\}, \\
\label{sta--insta-condi-stretch-2}
& {\rm unstable}, \quad {\rm if} \ (\lambda_1,\lambda_2) \in \mathcal S_{\rm unstable}=\left\{ (\lambda_1,\lambda_2) \in {\mathbb R}^+ \times {\mathbb R}^+: \frac{\textrm{max}\{\lambda_1,\lambda_2\}}{\lambda_1^2\lambda_2^2} \ge 3.383 \right\}.
\end{align}
\end{subequations}

\begin{figure}[t]
\centering
\subfigure[]{%
        \includegraphics[width=3in]{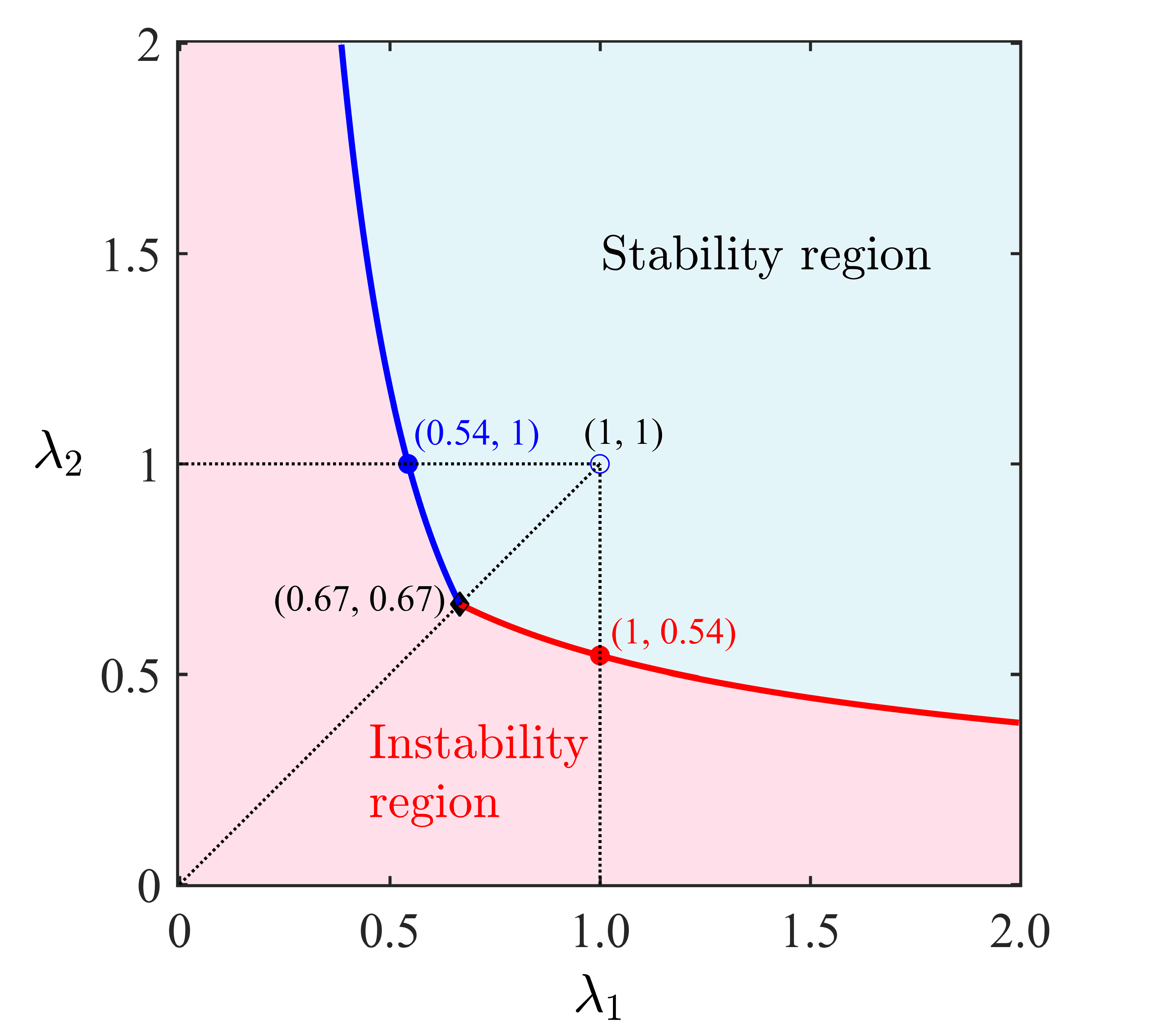}
       \label{fig4a} }
\subfigure[]{%
        \includegraphics[width=3in]{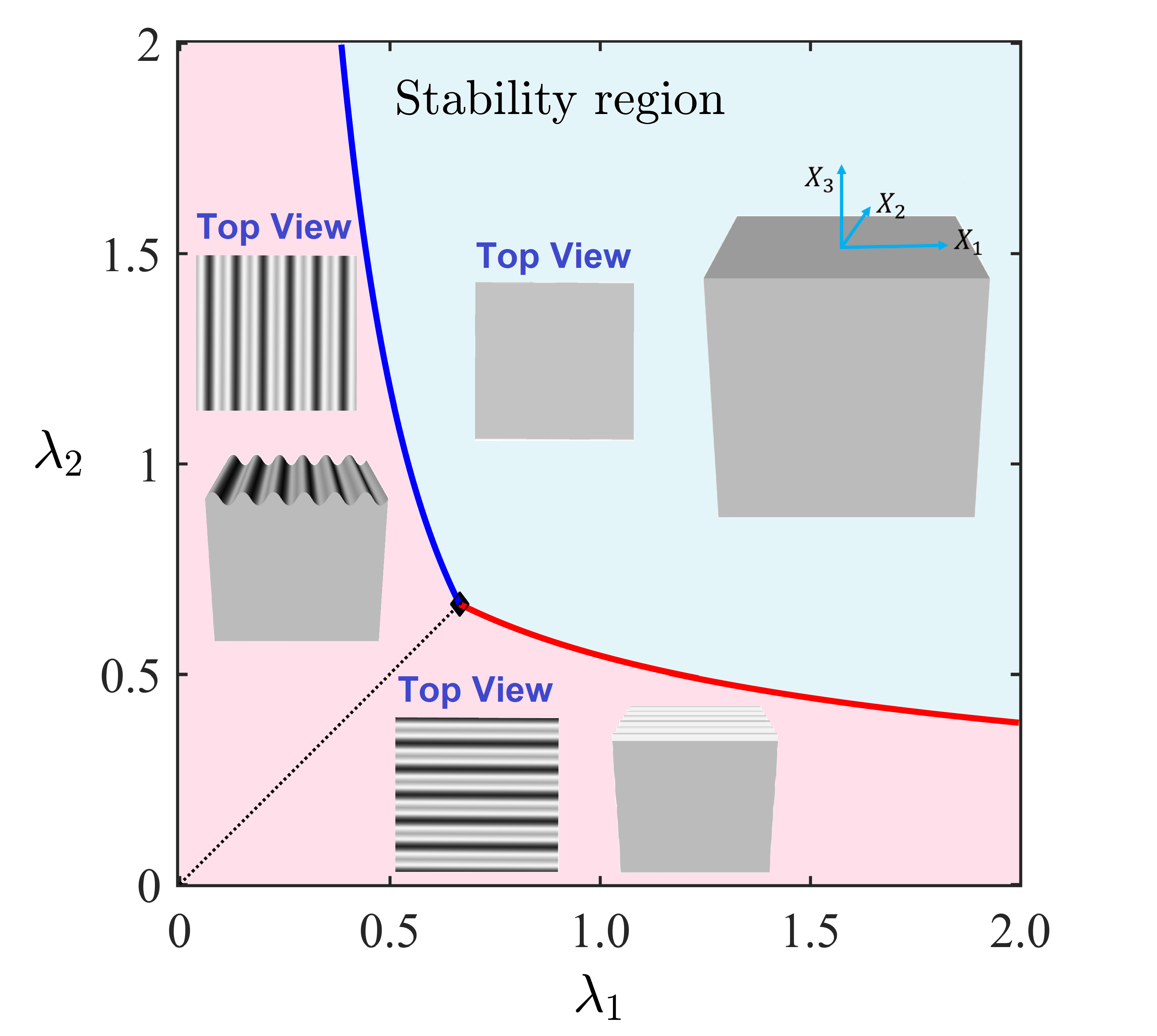}
       \label{fig4b} }
 \caption{(a) Stability and instability regions of surface wrinkling in the $\lambda_1-\lambda_2$ plane for a homogeneously deformed neo-Hookean block. (b) Surface patterns in the stability and instability regions. Wrinkles occur on the upper surface in the instability region while a flat upper surface is in the stability region. The wrinkle in the instability region is in the direction of a smaller principal stretch (larger compression).}
 \label{fig4stability}
 \end{figure}

Plot of the stability region $\mathcal S_{\rm stable}$ and the instability region $\mathcal S_{\rm unstable}$ in the $\lambda_1-\lambda_2$ plane is shown in Figure \ref{fig4stability}. In the following, detailed discussions are given for three particular cases (see Figure \ref{fig4a}): simple compression in the $X_1$ direction, compression in the $X_1$ direction under plane strain, and equi-biaxial compression the $X_1$ and $X_2$ directions.\\
\indent The first case admits free expansions in the $X_2$ and $X_3$ directions. The stretch $\lambda_1$, $0< \lambda_1< 1$, in the $X_1$ direction is less than $\lambda_2$ in the $X_2$ direction due to $\lambda_2=1/{\lambda_1}^{1/2}$. Then $\text{max}\{\lambda_1,\lambda_2\}$ is $\lambda_2=1/{\lambda_1}^{1/2}$ and the critical stretch from \eqref{sta--insta-condi-stretch} is obtained as $\lambda_1=(1/3.383)^{2/3}=0.444$ that is totally independent of the size of the block. This threshold $\lambda_1$ coincides with the critical stretch of surface wrinkling of an elastic half-space under plane strain (see eq.(4.14) in \cite{biot1963surface}). 

The second case corresponds to $\lambda_2=1>\lambda_1$ and then the critical stretch $\lambda_1$ from \eqref{sta--insta-condi-stretch} is given by $(1/3.383)^{1/2}=0.544$ (see Figure \ref{fig4a}), which is exactly the well-known \citeauthor{biot1963surface}'s prediction (see eq.(4.3) in \cite{biot1963surface}). Similarly, the third case of an equi-biaxial compression with $\lambda_1=\lambda_2$ achieves the critical value at $(1/3.383)^{1/3}=0.666$.

In spite of the coincidence, the deferences between this paper and Biot's work (\cite{biot1963surface}) may be noted. Biot's prediction is from the linear bifurcation analysis of a half-space under plane strain. The linear bifurcation analysis only gives the necessary conditions of the non-uniques solutions, however, it delivers little information about the stability and instability of the deformed body before and after the critical stretches. In addition, the surface instability problem studied by Biot is a two-dimensional rather than a three-dimensional analysis due to the plane strain assumption. In contrast to a half-space under plane strain, this paper presents a three-dimensional analysis of surface wrinkling of an elastic block subject to biaxial loading by an energy method. We directly show that the homogeneously deformed block has the lowest energy before the threshold and it is stable, however, it becomes unstable after the threshold since some states with lower energies have been found.

\subsection{Wrinkling patterns in the instability region in the $\lambda_1 - \lambda_2$ plane}
For any point $(\lambda_1, \lambda_2)$ in the instability region $\mathcal S_{\rm unstable}$ in Figure \ref{fig4stability}, the eigenvalue problem has numerous solutions of the eigenvalues $\mu$ and the corresponding eigenfunctions.

It is not hard to find that the stability conditions obtained from two-dimensional, for example the work \citep{biot1963surface, levinson1968stability}, and three-dimensional analyses seem the same. The maximum of $t_{mn}$ in $\eqref{t_mn_exp}$ is obtained at $m=0$ for $\lambda_1 > \lambda_2$ and at
$n=0$ for $\lambda_1 < \lambda_2$, which indicates directly that the equality in the stability condition $\eqref{sta--insta-condi-stretch}$ will be made firstly by a two-dimensional rather a three-dimensional perturbation (see Figure \ref{fig4stability}).

Surface instability will occur if $\frac{\textrm{max}\{\lambda_1,\lambda_2\}}{\lambda_1^2\lambda_2^2}$ slightly exceeds the threshold value of $3.383$ for some stretches in $\eqref{sta--insta-condi-stretch}$ and instability is more likely to happen for two-dimensional perturbations that correspond the maximum in $\eqref{t_mn_exp}$. However, the stability of large effective wavelengths will be influenced by the height $l_3$ of the rectangular block. Since a larger wavelength corresponds to a smaller dimensionless wavenumber $\bar k_{mn}=k_{mn} l_3$ at a given height $l_3$, then a small $\bar k_{mn}$ tends to require a smaller $t$ for surface instability (see Figure \ref{fig2b} or Figure \ref{fig3}). Physically, the prescribed displacement $\eqref{RB_C_2}$ at the bottom surface $X_3 = l_3$ tends to stabilize perturbations with large effective wavelength. As a consequence, the critical stretches of surface instability are independent of the block size but the wrinkled patterns only appear for
wavelengths which are smaller than the threshold that depends on the height of the rectangular block.

\section{Conclusions} \label{conclusion}
Motivated by the fantastic surface patterns and a variety of applications by harnessing surface instabilities recently, we study the surface wrinkling of a finite block of elastic materials subject to biaxial loading by an energy method. In contrast to the linear stability analysis, the energy method can give a complete set of stability conditions including the stability and instability regions. We perform the first and second variations of the energy functional. The second variation condition is transformed into an eigenvalue problem that is solved by using double Fourier series. The requirement of all nonnegative eigenvalues gives the stability condition and the comparison of the energy provides the wrinkle patterns in the instability region. We briefly summarize our results below:

1. The homogeneously deformed block (flat surface) is stable before the principal stretches reach the threshold. These regions can only be obtained by using the energy method rather the linear stability analysis.

2. The boundary between the stability (flat surface) and instability (surface wrinkling) regions in the principal stretch plane is independent of the size of the block. 

3. The boundary coincides with the threshold of the surface instability of an elastic half-space problem that is studied by \cite{biot1963surface} with the linear stability analysis. The coincidence implies some relations between surface instabilities of the homogeneous deformation of elastic bodies with finite and infinite domains. However, the general relation between the surface wrinkling and the boundary conditions as well as the geometry of the elastic body is still an open question.

4. In the instability region the surface wrinkling appears in the direction of the smaller stretch (the higher compressive strain). In other words, a two-dimensional perturbation has lower energy and is more natural to trigger the surface wrinkling. 

5. Surface instability only appears for perturbations of small wavelengths restricted by the block height. For an infinite height, all the wavelengths of two-dimensional perturbations become unstable in the instability region.
 
We hope our study of surface wrinkling of a finite block of elastic materials could shed light on the relation of surface instabilities between finite and infinite bodies. The energy method may broaden our horizons of the fundamental issues of stability and bifurcation in the topic of surface instabilities.

\section*{Acknowledgements} 
The author wishes to express his deepest gratitude to Professor Yi-chao Chen at the University of Houston for helpful comments on the manuscript.

\end{document}